\documentclass[reprint,showpacs,amsmath,amssymb,aps,pre]{revtex4-1}

\usepackage{dcolumn}
\usepackage{bm}
\usepackage{amsmath}
\usepackage{amssymb}
\usepackage{graphicx} 
\usepackage{csquotes}
\usepackage[nooneline, tight,raggedright,FIGTOPCAP]{subfigure}
\usepackage[linkcolor=black,citecolor=blue,urlcolor=blue,colorlinks=true]{hyperref}
\usepackage[open, openlevel=0]{bookmark}
\usepackage{tabularx}
\usepackage{tabulary}
\usepackage{chemarr}
\usepackage[normalem]{ulem}
\usepackage{color}
\usepackage{epsfig}
\usepackage{enumerate}

\newcommand{\bequ}{\begin{equation}}
\newcommand{\eequ}{\end{equation}}

\newcommand{\dt}{\mathrm{d}t}
\renewcommand{\d}{\mathrm{d}}

\newcommand{\crossingarrows}{\begin{picture}(5,5) \put(0,0){$\nearrow$} \put(0,0){$\nwarrow$} \end{picture} }
\renewcommand{\rightleftharpoons}{\begin{picture}(5,5) \put(0,3){$\rightarrow$} \put(0,0){$\leftarrow$} \end{picture} }
\newcommand{\labfig}[1]{\label{fig:#1}}

\newcommand{\labsec}[1]{\label{sec:#1}}
\newcommand{\reffig}[1]{\ref{fig:#1}}

\newcommand{\refsec}[1]{\ref{sec:#1}}
\newcommand{\refsfig}[1]{\subref{fig:#1}}
\newcommand{\circled}[1]{\protect\raisebox{.5pt}{\textcircled{\protect\raisebox{-1pt} {#1}}}}
\definecolor{darkgreen}{rgb}{0,0.5,0} 
\definecolor{violet}{rgb}{0.5,0,0.5}
\definecolor{orange}{rgb}{0.2,0.5,0.5}

\newcommand{\resug}[1]{{\color{red}#1}}
\newcommand{\rest}[1]{{\footnotesize \color{red}\xout{#1}}}
\newcommand{\edsug}[1]{{\color{magenta}#1}}
\newcommand{\edst}[1]{{\footnotesize \color{magenta}\xout{#1}}}
\newcommand{\authsug}[1]{{\color{green}#1}}
\newcommand{\authst}[1]{{\footnotesize \color{green}\xout{#1}}}
\newcommand{\changed}[1]{{\color{red}#1}}

\renewcommand{\rest}[1]{}
\renewcommand{\resug}[1]{{#1}}
\renewcommand{\edst}[1]{}
\renewcommand{\edsug}[1]{{#1}}
\renewcommand{\authst}[1]{}
\renewcommand{\authsug}[1]{{#1}}
\renewcommand{\changed}[1]{}

\begin{document}  
\title{Cooperative effects enhance the transport properties of molecular spider teams}
\author{Matthias Rank}
\author{Louis Reese}
\author{Erwin Frey}
\email{frey@lmu.de}

\affiliation{Arnold Sommerfeld Center for Theoretical Physics (ASC) and Center for NanoScience (CeNS), Department of Physics, Ludwig-Maximilians-Universit\"at M\"unchen, Theresienstra\ss e 37, 80333 M\"unchen, Germany}

\begin{abstract}
Molecular spiders are synthetic molecular motors based on DNA na\-no\-technology.
While natural molecular motors have evolved towards very high efficiency, it remains a major challenge to develop efficient designs for man-made molecular motors.
Inspired by biological motor proteins like kinesin and myosin, molecular spiders comprise a body and several legs. The legs walk on a lattice that is coated with substrate which can be cleaved catalytically.
We propose a novel molecular spider design in which $n$ spiders form a team.
Our theoretical considerations show that coupling several spiders together alters the dynamics of the resulting team significantly. 
Although spiders operate at a scale where diffusion is dominant, spider teams can be tuned to behave nearly ballistic, which results in fast and predictable motion.
Based on the separation of time scales of substrate and product dwell times, we develop a theory which utilises equivalence classes to coarse-grain the micro-state space. 
In addition, we calculate diffusion coefficients of the spider teams, employing a mapping of an $n$-spider team to an $n$-dimensional random walker on a confined lattice. 
We validate these results with Monte Carlo simulations and predict optimal parameters of the molecular spider team architecture which makes their motion most directed and maximally predictable.
\end{abstract}

\date{\today}

\maketitle

\section{Introduction}
How the motion of molecules along predefined traffic routes emerges and how these molecules self-organise is now an experimentally tractable question due to advances in nanotechnology. Molecular motors that have evolved inside cells and perform well-defined tasks~\cite{Howard2001} inspired the engineering of DNA devices performing motor business on the nanoscale~\cite{Bath2007, Astumian2010, Pinheiro2011}: so-called DNA walkers have been build that move or diffuse along a substrate~\cite{Sherman2004, Shin2004, VonDelius2011}. Among the first autonomous synthetic walkers was a motor design that used a catalytic reaction to cleave a substrate in order to move forward~\cite{Tian2005}. Since then, a plethora of different motor molecules have been built from scratch in the laboratory. They do not only serve technological advances, but also shed light on the basic principles of molecular movement, e.g. of biological molecular motors. One class of molecules that attracted a great deal of attention are molecular spiders~\cite{Pei2006}. They combine the catalytic activity of nucleic acids with a multivalent design: attached to a body are several legs of single-stranded DNA. These DNA legs can bind to and catalytically cleave a substrate. This can be repeated over and over again, which in turn generates processive motion: while individual legs dissociate from the substrate on a timescale of seconds, the multipedal architecture ensures tight binding of the spider to the substrate for hours~\cite{Pei2006}. Recent experiments used DNA origami to build quasi-one-dimensional tracks for molecular spiders~\cite{Lund2010}. A predescribed substrate landscape allows to assign special tasks to a spider and for instance control its movement. The simple yet well-defined design makes it possible to study spiders in great detail and probe theoretical predictions.

Molecular spiders have also been theoretically studied extensively in recent years. Antal et al.~\cite{Antal2007a} and Antal and Krapivsky~\cite{Antal2007} were the first to propose an abstract model that describes the dynamics of molecular spiders. They analysed the spiders' kinetics for various architectures and found a variety of interesting effects which arise due to the mutual exclusion of spider legs on the lattice and the presence of the substrate. Substrates are cleaved \resug{slowly} \rest{at rate $r<1$ which is slow} in comparison to hopping from already cleaved sites. This distinction leads to subtle memory effects that affect the spiders dynamics and result in a bias towards the substrate~\cite{Antal2007}. When the spider is in an all-cleaved area, principles emerging from simple exclusion processes~\cite{Chou2011, Mobilia2008}  allow a derivation of the spiders' diffusion constants~\cite{Derrida1995}.

In the meantime, mechanistically more detailed systems have been considered. These include the variation of the rate constants involved in the chemical reactions~\cite{Semenov2011,Samii2010}  and boundary conditions~\cite{Samii2010}, as well as the number and length of legs~\cite{Samii2011}. Samii et al.~\cite{Samii2010} investigated  the spiders' stepping gait and considered inchworm as well as hand-over-hand spiders. Semenov et al.~\cite{Semenov2011} showed that spiders experience a rather extended time period of superdiffusion given that the cleavage rate $r$ is small. More complex spiders in quasi-one~\cite{Olah2012} and in two dimensions~\cite{Antal2012} have also been studied. Moreover, there have also been recent studies focussing on mathematical aspects like recurrence, transience and ergodicity~\cite{Gallesco2011, Ben-Ari2011}, as well as random environments~\cite{Gallesco2011a, Juhasz2007}. These investigations have examined molecular spiders independently from their chemical motivation as a general class of multivalent random walkers~\cite{Olah2012}.

The rich variety and diversity of these recent studies show that molecular spiders are a versatile system to study artificial molecular motors both theoretically as well as experimentally. However, many challenges still remain in improving their efficiency and tailoring the spiders' design for possible biotechnological applications~\cite{VonDelius2011}.

In this study, we examine dynamic and stochastic properties of a novel molecular spider design: $n$ molecular spiders are constrained due to their joint attachment to a single linking node which may be considered as a primitive model of a cargo. The resulting spider-spider interactions lead to collective effects which enhance the motor properties of the $n$-spider team. We show that spider teams are faster and move more persistently along their track than individual spiders.
We also predict that the spider teams move at reduced randomness and thus are candidates for applications that require reliable, \emph{i.e.} predictable motion~\cite{Pinheiro2011}. 

This paper is organised as follows: in Sec.~\refsec{model} we provide a detailed picture of how molecular spiders function and give a comprehensive introduction to the existing theoretical models before we define the dynamics of an $n$-spider team. Subsequently, in Sec.~\refsec{sec_enhanced_properties} we present our main results: spider teams have enhanced motor properties. To explain these numerical observations, we give a comprehensive analysis of the stochastic dynamics of a spider team. In particular, we perform a reduction of the state space of the spider teams and thereby calculate the mean number of consecutive directed steps a spider team performs while attached to the substrate boundary (Sec.~\refsec{sec_boundary_periods}). Moreover, we explore the validity of the resulting network representation of the spider team dynamics and also show how it breaks down (Sec.~\refsec{sec_validity_eq_classes}).
In addition to this approach, we provide an exact mapping of the $n$-spider team to an $n$-dimensional confined random walk (Sec.~\refsec{sec_mapping_conf_rw}). This enables us to quantify the diffusion coefficient which describes the motion of a spider team during diffusive periods (Sec.~\refsec{sec_diffusive_periods}).
Finally, in Sec.~\refsec{optimisation} we bridge theoretical and experimental observables and predict the existence of optimal parameters which maximise the spider teams' predictability. Finally (Sec.~\refsec{conclusions}), we conclude and identify connections to related fields.

\section{Model definition}\labsec{model}
\begin{figure}
\centering
\subfigure[\labfig{fig_model_sketch}]{\includegraphics[width=4.5cm]{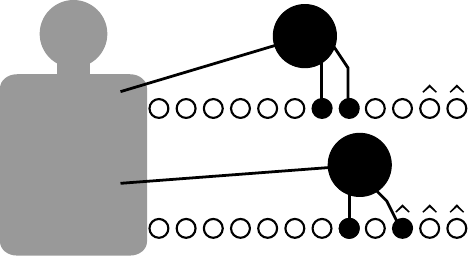}}
$\qquad$\subfigure[\labfig{fig_d_sketch}]{\includegraphics[width=2.5cm]{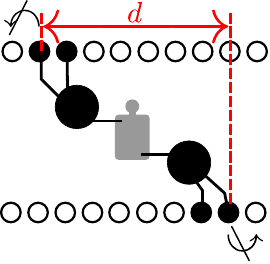}}
\caption{\edsug{(Colour online)} Cartoon of the spider team model and definition of the leash length $d$. \subref{fig:fig_model_sketch} \changed{[Changed configuration of the lower spider]} Two spiders are attached to a joint cargo with an inelastic string. Both spiders walk on their respective one-dimensional track. Hats indicate the presence of substrate. \subref{fig:fig_d_sketch} The finite length of the linking string induces a maximal distance between the spiders' bodies which gives rise to a maximal span of the spider team, characterised by the \enquote{leash length} $d$.\labfig{model}}
\end{figure}

Our model is based on the theoretical description of molecular spiders introduced by Antal et al.~\cite{Antal2007a} and Antal and Krapivsky~\cite{Antal2007} that was motivated by experiments of Pei et al.~\cite{Pei2006}. 
They propose a spider design that consists of a central body and $l$ legs that are attached to it.
Each leg has a certain length and thus the overall spider can span a maximal distance $s$.
In the experiment, a spider is exposed to a (one-dimensional) lattice, to which a substrate is attached. Since binding of leg and substrate happens through the Watson-Crick mechanism~\cite{Watson1953}, only one leg may bind to a lattice site at a time.
In the model, this corresponds to an exclusion process in that the movement of one spider leg is constrained by the spider's remaining legs.
The lattice prevails in two states: with and without substrate. Legs which bind to lattice sites with substrate can remove it (chemically: they cleave it, only a shorter part remains bound to the lattice), which happens along with unbinding from that site at rate $r$. By contrast, spiders unbind from sites without substrate (\emph{i.e.} from product sites) at rate~$1$. In the model, a substrate is \emph{always} cleaved when a leg steps away from it, and rebinding of a leg to a new lattice site happens instantaneously. 
Two different rules to rebind to a new lattice site have to be distinguished: spiders' legs either have a certain ordering, \emph{i.e.} they cannot \enquote{overtake} each other; these spiders are termed \emph{inchworm} spiders~\cite{Antal2007,Antal2007a,Semenov2011}. Alternatively, spider legs have no ordering, they can step over each other; those spiders have been called \enquote{quick spiders}~\cite{Antal2007a} or \enquote{hand-over-hand} spiders~\cite{Samii2010,Samii2011} in previous studies. 
Both types of spiders show quite different behaviour~\cite{Samii2011} and have to be well distinguished. In this paper, we will concentrate on inchworm spiders. 

Although in our model a leg which has just unbound from the lattice rebinds to the lattice instantaneously, we allow a spider's leg to rebind to any lattice site as long as the new leg configuration does not violate any of the restrictions imposed by the leg length or the ordering of the legs. In particular, this implies that rebinding to the lattice site from which the leg just unbound is possible~\footnote{This differs from the original model of Antal et al.~\cite{Antal2007a} who allowed rebinding only to \emph{different} sites.}; this can be motivated from experiments where the typical timescales for binding to substrates exceed those for diffusion by orders of magnitude~\cite{Samii2010}. In addition, our choice obviates unphysical situations that might occur for spider teams due to the complete blockage of a leg.

Hollow circles ($\circ$) denote unoccupied lattice sites, filled circles ($\bullet$) indicate that a leg is attached to that site. The presence of substrate is marked with a hat, \emph{i.e.} $\hat \bullet$, or $\hat \circ$. Throughout this paper, we consider bipedal spiders (\emph{i.e.} $l=2$) with a maximal leg span of $s=2$. Spiders may thus only arise in either the spanned ($\arraycolsep.5pt \begin{array}{ccc} \bullet & \circ & \bullet \end{array}$) or the relaxed ($\arraycolsep.5pt \begin{array}{cc} \bullet & \bullet \end{array}$) configuration. For this case, the geometry of the cleaved sites, which is usually called \emph{product sea}, is an interval on the one-dimensional lattice; it gives rise to memory effects which stem from irreversible substrate cleavage~\cite{Antal2007a}.

Samii et al.~\cite{Samii2011} suggested that the lattice could be prepared with substrates on the right, and products on the left hand side from the very beginning, and called this initial condition \emph{P-S lattice}. This asymmetry makes some calculations easier, and it provides a symmetry breaking direction already at the beginning of the dynamics. We are going to use this kind of lattice throughout this paper.

\begin{figure*}
\subfigure[]{\includegraphics[width=0.315\textwidth]{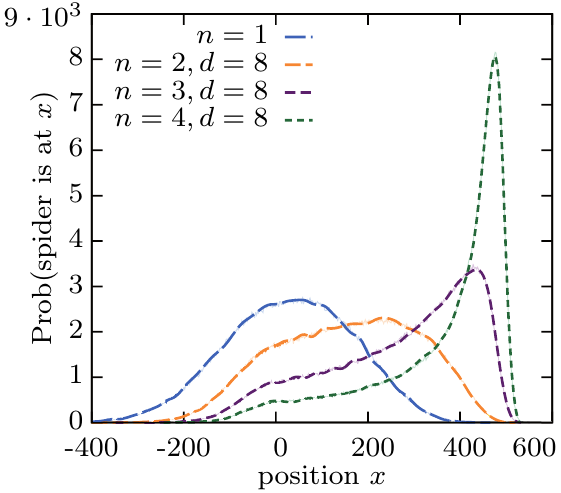}\labfig{fig_prob_distr}}
$\quad$\subfigure[]{\includegraphics[width=0.315\textwidth]{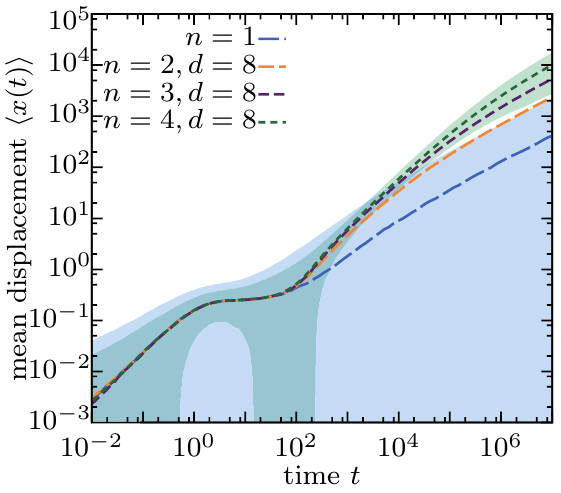}\labfig{fig_mean_displacement}}
$\quad$\subfigure[]{\includegraphics[width=0.315\textwidth]{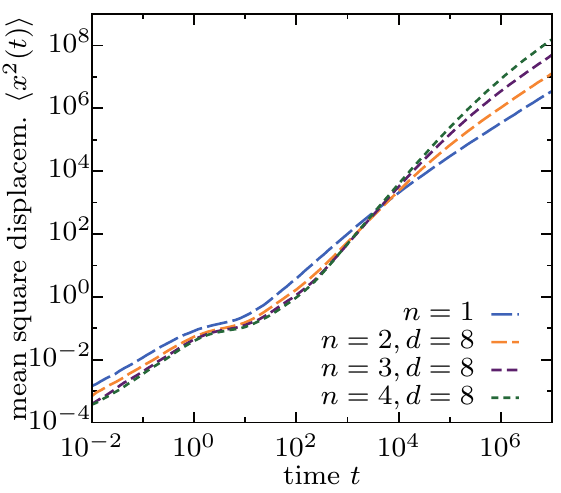}\labfig{fig_mean_square_displacement}}
\subfigure[]{\includegraphics[width=0.315\textwidth]{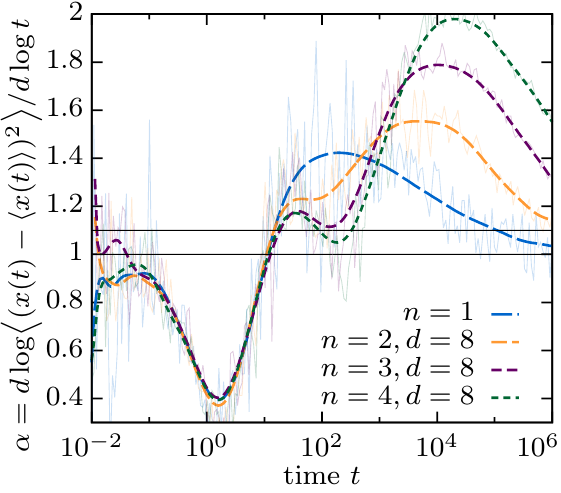}\labfig{fig_plot_alpha}}
$\quad$\subfigure[]{\includegraphics[width=0.315\textwidth]{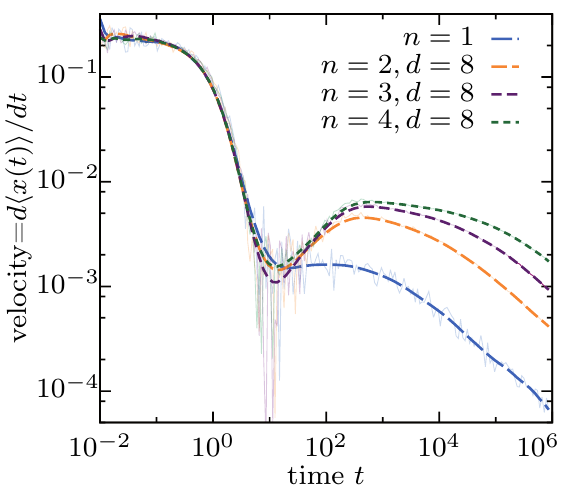}\labfig{plot_velocity}}
$\quad$\subfigure[]{\begin{minipage}{.315\textwidth}\includegraphics[width=\textwidth]{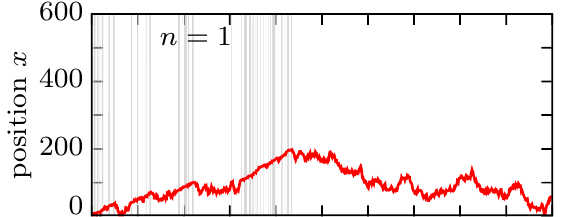} 
\includegraphics[width=\textwidth]{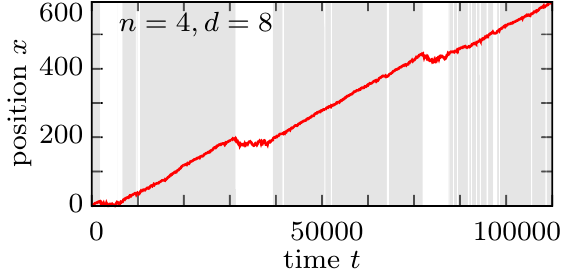}
\end{minipage} \labfig{fig_sample_trajectory}}
\caption{\changed{[Merged figures (was originally two figures (a)-(c) and (d)-(e))]}(Colour online) Dynamic properties of spider teams. \edsug{Positions are given in lattice units throughout this work; time is defined by setting the hopping rate from products to $1$.}  Thin shaded lines show data from finite difference approximations, thick lines show smoothing Bezier curves.\refsfig{fig_prob_distr} Probability distributions (histograms) of spiders to be at position $x$ at time $t=10^6$; simulation data were binned with a box size $1$.  
Depicted are distributions for a single spider and spider teams with $n=2, 3,4$ and $d=8$, and cleavage rate $r=0.01$. While the single spider distribution follows nearly a Gaussian centred close to the origin, the distributions of spider teams are clearly skewed and shifted towards larger $x$. 
The asymmetry stems from the P-S preparation of the lattice at $t=0$ (products at the left, substrates at the right)~\cite{Samii2010}.
\refsfig{fig_mean_displacement} Mean displacement as a function of time (lines). The shaded areas represent the standard deviation around the mean displacement for a single spider and the $4$-spider team, respectively, and provide a measure for the randomness of the spiders' motion.
Note that the \emph{visual} impression of the standard deviation is rather that of a relative deviation, since the plot is in double logarithmic scale. \resug{For a discussion of the temporal behaviour of the dynamic quantities see the main text.}
\refsfig{fig_mean_square_displacement} Mean square displacement (MSD) as a function of time, $\langle x^2(t)\rangle$. \authst{For a more detailed analysis of the MSD's effective exponent see Fig.~\reffig{speed}\refsfig{fig_plot_alpha}.}
\labfig{properties} \authst{\refsfig{fig_plot_alpha}-\refsfig{plot_velocity}: Effective exponent of the MSD and velocity of spider teams (cleavage rate $r=0.01$).} \refsfig{fig_plot_alpha} \changed{[Changed figure; now using a new definition of $\alpha$.]} The \authst{MSD's} \authsug{variance's} effective exponent $\alpha(t)$, see Eq.~\eqref{eq_alpha}. For diffusion, $\authst{\langle x^2(t) \rangle} \authsug{ \bigl\langle (x(t)-\langle x(t) \rangle)^2 \bigr\rangle} \propto t^1$, hence $\alpha=1$; superdiffusion corresponds to $\alpha>1.1$~\cite{Semenov2011}, and ballistic motion to $\alpha=2$. 
The superdiffusive regime of spider teams lasts longer than that of single spiders; large spider teams reach nearly-ballistic motion for significantly long times. 
\refsfig{plot_velocity} Mean velocity of the spiders as a function of time. The mean velocity is defined as the time-derivative of the mean displacement, $d\langle x (t) \rangle/dt$. Spider teams outperform single spiders by an order of magnitude. \refsfig{fig_sample_trajectory} Sample trajectory of a single spider (top), and a $4$-spider team with $d=8$ (bottom). Periods in which the spider (team) is in the vicinity of the product-substrate boundary are shaded.
\labfig{speed}}
\end{figure*}

Taken together, the spiders which we examine in this study are bipedal ($l=2$) inchworm spiders with a maximal span of $s=2$, which walk on a one-dimensional P-S lattice. Every spider's leg may rebind to any accessible lattice site as long as the ordering is preserved, including the site from where it just unbound. 

Based on this model for molecular spiders, we propose a minimal model for a team of molecular spiders.
Several, say $n$, molecular spiders are linked to a (virtual) cargo with an inelastic leash (\emph{i.e.} a string) of a well-defined length $\tilde d$. Each of these spiders runs on its own one-dimensional track\rest{,}\resug{. This is} similar to biological molecular motors like kinesin-1~\cite{Brunnbauer2012,Ray1993} that walk along \resug{one-dimensional} microtubule filaments~\cite{Howard2001}. We call these ensembles of spiders that jointly pull a cargo a spider team. For a cartoon of a team of two spiders, see Fig.~\reffig{model}.

Note that the role of the \enquote{cargo} is not primarily to put load on the spiders, actually we set the mass of the cargo equal to zero. In contrast, the cargo mediates the interaction among the $n$ spiders comprising the team: 
since the strings used for linking the spiders to the cargo are inelastic with some length $a$, any two of the spiders' bodies may mostly be $2 a$ away from each other. From the bodies, the furthestmost reachable lattice site is given by the spiders' legs' length, call it $b$, so that the \emph{maximal} distance between the leftmost and the rightmost leg of all the spiders in the team is given by $2(a + b) =:d$.   Mathematically, letting $\lambda_i$ [$\rho_i$] denote the position of the $i$th spider's left [right] leg, this restriction reads \bequ \vert \rho_i-\lambda_j \vert \leq d ~ \forall i,j \, . \eequ
Note that this is a global constraint which restricts the spider \emph{team}, in contrast to the local constraint limiting the span of an \emph{individual} spider, 
\bequ \vert \rho_i-\lambda_i \vert \leq s ~ \equiv 2 ~ \forall i. \eequ
The definition of $d$ is visualised for a 2-spider team in Fig.~\reffig{model}\refsfig{fig_d_sketch}.
For simplicity of language, and to capture an intuitive understanding especially for $2$-spider teams, we will call $d$ the \emph{leash length} in the following.

\section{Results\labsec{results}}
\subsection{Enhanced properties of $n$-spider teams\labsec{sec_enhanced_properties}}
\labsec{prop}

We performed extensive numerical simulations to characterise the dynamic properties of $n$-spider teams. Our simulation data show that the constraint arising through the leash that holds the spider team together induces collective effects among the $n$ spiders. We find that the incorporation of a spider into a team enhances many of the motor properties: the mean travelled distance of a spider team exceeds that of single spiders by far, up to orders of magnitude, for a rather small cleavage rate $r=0.01$, see Figs.~\reffig{properties}\refsfig{fig_prob_distr},\refsfig{fig_mean_displacement}. In addition, a spider team's movement is a lot more \enquote{predictable}. This can be inferred from the width of the probability distributions, see Fig.~\reffig{properties}\refsfig{fig_prob_distr}, and the shaded areas depicted in Fig.~\reffig{properties}\refsfig{fig_mean_displacement}, which illustrate the standard deviation of the mean displacement.

Another important quantity is the mean square displacement (MSD) of the spider teams, see Fig.~\reffig{properties}\refsfig{fig_mean_square_displacement}. It shows a steep increase at intermediate timescales, similar but stronger and longer-lasting compared to recent results by Semenov et al.~\cite{Semenov2011} for single spiders: in this regime spiders move superdiffusively. To quantify the time-dependent effects of superdiffusion, we evaluated the \enquote{slope} of the \authst{MSD in a double logarithmic scaling like in Fig.~\reffig{properties}\refsfig{fig_mean_square_displacement}} \authsug{variance in a double logarithmic scaling}, \emph{i.e.} the effective exponent 
\bequ 
\alpha(t)=\frac{\authst{d \log \langle x^2(t) \rangle }}{\authst{d \log t }}\authsug{=\frac{d \log \bigl\langle (x(t)-\langle x(t) \rangle)^2\bigr \rangle}{d \log t}}
\label{eq_alpha}\, ;
\eequ 
which provides a measure for diffusivity, see also Ref\authsug{s}.~\authsug{\cite{Semenov2011,Olah2012,Bouchaud1990}}. Figure~\reffig{speed}\refsfig{fig_plot_alpha} shows $\alpha(t)$ for a single spider and several different spider teams. Remarkably, the 4-spider team travels almost ballistically and the periods of \enquote{instantaneous superdiffusion} of spider teams (\emph{i.e.} times with $\alpha > 1.1$~\cite{Semenov2011}) last much longer compared to single spiders. The nontrivial shape of $\alpha(t)$ indicates the multitude of dynamic processes that are involved in the spider team's dynamics\rest{.}\resug{: initially, $\alpha \approx 1$ for $t \lesssim 1$ for all configurations, reflecting the very first hop of the spiders' left legs. In succession, until $t \lesssim r^{-1}=100$, the spiders' right legs have typically not yet cleaved a substrate, whereas the left legs jump back and forth, hence the variance is approximately constant and thus $\alpha<1$ (for these two regimes, see also a more explicit discussion in Ref.~\cite{Semenov2011}). Had we chosen other starting conditions for the spiders, the behaviour at short time would look different. 
Likewise, also the following regime until $t \lesssim 10^2 \dots 10^3$ results from the fixed starting conditions: while at early times the spider team does not feel the leash and all spiders can move independent from each other, at some point the leash is fully spanned and the spiders at the most extreme position (\emph{i.e.} those contributing most to the variance) are retarded. This leads to a transient decrease of $\alpha$. This regime is unique to spider teams since it is an effect constituted by the leash. Finally, for large times $t \gtrsim 10^2 \dots 10^3$, the memory of initial conditions is lost and $\alpha$ becomes maximal. Clearly, the maximal value of $\alpha$ is greatest for $n=4$ of the displayed configuration. As time increases further, $\alpha$ decreases slowly which is due to the fact that more and more spiders move away from the product-substrate boundary (see also Ref.~\cite{Semenov2011}).
}
Figure~\reffig{speed}\refsfig{plot_velocity} shows the velocity of the spider team by means of the derivative of the mean displacement with respect to time. Clearly, the velocity of a 4-spider team outperforms that of a single spider by more than one order of magnitude.

These pronounced effects are in a way surprising: at first sight, one might speculate that the coupling leash which imposes an additional constraint on the spiders would handicap the spider team's motion and make it slower. This is clearly not the case.  To the contrary, the dynamic properties of the spider teams are enhanced.
In the remainder of this section we will explain this effect using analytical arguments.

\subsection{Boundary periods\labsec{sec_boundary_periods}}

\begin{figure*}
\subfigure[\labfig{fig_boundary_period_single}]{\includegraphics[width=6.5cm]{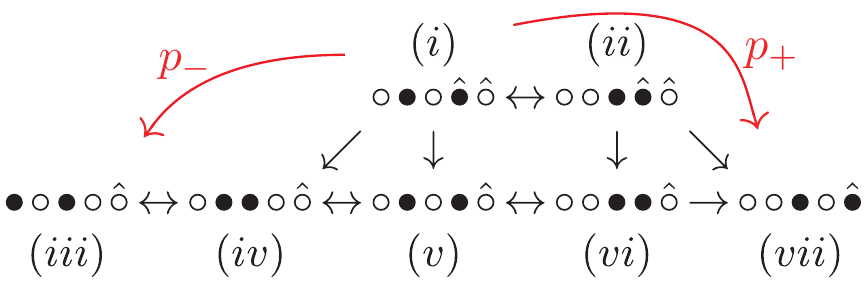}} ~
\subfigure[\labfig{fig_boundary_period_team}]{\includegraphics[width=9.5cm]{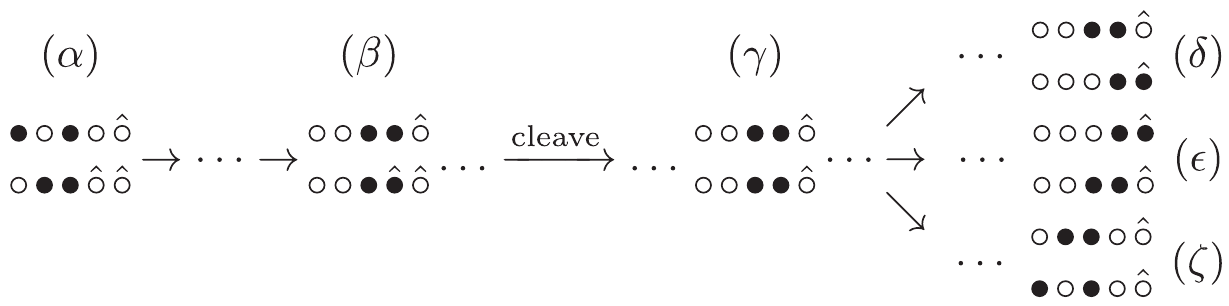}}
\caption{\edsug{(Colour online)} Definition of a boundary period. \refsfig{fig_boundary_period_single} Path of a single spider through a boundary period. The period \emph{always} starts in state $(i)$. From there, the spider can change to $(ii)$, and back. When the right leg cleaves the substrate, the spider arrives at $(iv), (v), (vi)$, or $(vii)$. Arriving at $(vii)$ corresponds to continuing the same boundary period from a new substrate (with \enquote{$(vii)$ being the new $(i)$}), since $(vii)$ and $(i)$ are equivalent up to translation. Hence, the number of steps is raised by $1$ upon arriving at $(vii)$. If, by contrast, the spider reaches $(iii)$, the boundary period ends and a diffusive period begins. The probability to make a successful step, \emph{i.e.} to reach $(vii)$ before $(iii)$, is the bias $p_+$ calculated by Antal and Krapivsky~\cite{Antal2007}. The number of steps during a boundary period is then the number of transitions $(i) \to (vii)$, without reaching $(iii)$ in between. This is equivalent to the number of cleavages during a boundary period, not counting the very last cleavage (which is not counted since by definition the spider steps away from the boundary after the last cleavage, and we only count forward steps).
\refsfig{fig_boundary_period_team} Example of a boundary period of a $2$-spider team. $(\alpha)$ None of the spiders is in a boundary period, hence none of them experiences a bias. Thus, the spider \emph{team} is in a diffusive period. When the lower spider reaches a substrate $(\beta)$ it enters a boundary period. Thus, also the spider \emph{team} enters a boundary period. In succession, the lower spider's right leg happens to cleave the substrate $(\gamma)$. The lower spider can then find its way to a new substrate $(\delta)$ what constitutes $\frac 12$ successful step for the spider \emph{team} and preserves the boundary period. If the upper spider, in this case, steps to a substrate $(\epsilon)$, this does, however, not yet constitute a step. 
This is because although the spider \emph{team} is in a boundary period, the 
\emph{upper} spider has not been in a boundary period itself during this team's boundary period. Since a step essentially reflects a cleavage, no step can be integrated in this case.
If the lower spider steps away from the new substrate $(\zeta)$, the spider team enters a diffusive period. 
In analogy to single spiders, the number of steps during a spider team's boundary period is equivalent to the number of cleavages during that period, divided by the number of spiders, and not counting each spider's \emph{last} cleavage event.
\labfig{fig_boundary_periods}
}
\end{figure*}

\paragraph{Single spiders.} 
Key to the understanding of an individual molecular spider's motion is to unravel the mechanism for biased motion. To this end we distinguish between two qualitatively different dynamic states of the spiders: looking at single trajectories of molecular spiders we find that there are periods of time in which the spider's motion is strongly directed, and other periods with undirected, diffusive motion (see Fig.~\reffig{speed}\refsfig{fig_sample_trajectory}). In the following, we will call these dynamic states \emph{boundary periods} and \emph{diffusive periods}, respectively. To define the notion of these periods, it is convenient to distinguish between the steps of the spider's legs and the step of the spider as a whole. We define a \emph{spider step} as a transition from a spread configuration ($\arraycolsep.5pt \begin{array}{ccccc} \circ & \bullet & \circ & \bullet & \circ \end{array}$) to another spread configuration shifted by one lattice unit forwards or backwards, \emph{i.e.} $\arraycolsep.5pt \begin{array}{ccccc} \circ & \circ & \bullet & \circ & \bullet \end{array}$ or $\arraycolsep.5pt \begin{array}{ccccc} \bullet & \circ & \bullet & \circ & \circ \end{array}$, irrespective of the sites being products or substrates. During a diffusive period all the spider's legs are attached to product sites and therefore the spider steps with equal probability in both directions~\cite{Antal2007a}. In contrast, biased spider motion can emerge in the vicinity of the boundary between product and substrate sites. We define a \emph{boundary period} as follows: it starts with a spread configuration where the right spider leg is attached to a substrate ($\arraycolsep.5pt \cdots \begin{array}{ccccc} \circ & \bullet & \circ & \hat \bullet & \hat \circ \end{array} \cdots $), and ends when the spider has fully stepped away from the substrate boundary ($\arraycolsep.5pt \cdots \begin{array}{ccccc} \bullet & \circ & \bullet & \circ & \hat \circ \end{array} \cdots $), (the dots indicate that the block of displayed lattice sites may have been shifted during the boundary period) as illustrated in Fig.~\reffig{fig_boundary_periods}\refsfig{fig_boundary_period_single}. As a consequence, during a boundary period the substrate boundary is shifted by an integer number of lattice units forward.

For single spiders the bias can be measured by calculating the \emph{first passage probability}, $p_+$, for the spider to progress one step forward during a boundary period, \emph{i.e.} $p_+=$Prob$\lbrace \arraycolsep.5pt \begin{array}{ccccc}\bullet & \circ & \bullet & \circ & \hat \circ \end{array} \not \leftarrow \begin{array}{ccccc} \circ & \bullet & \circ & \hat \bullet & \hat \circ \end{array} \rightarrow \arraycolsep.5pt \begin{array}{ccccc} \circ & \circ & \bullet & \circ & \hat \bullet \end{array} \rbrace $; see also Fig.~\reffig{fig_boundary_periods}\refsfig{fig_boundary_period_single} for an illustration of the corresponding dynamic processes.
By analysing all possible sequences of transitions, Antal and Krapivsky found an explicit expression for the bias, namely $p_+(r)=\frac{5+r}{8+4r}$~\cite{Antal2007}, valid for spiders with legs always jumping to \emph{neighbouring} sites. Similar calculations can be performed for spiders whose legs may also rebind to the \emph{same} site again (like those we consider throughout this paper), leading to $\tilde{p}_+(r)=\frac{5+3r}{8+8r}$. The mathematical expressions for $p_+$ and $\tilde p_+$ differ only slightly, in particular they are equal in the limits $\lim_{r \rightarrow 0} \tilde{p}_+(r)= \lim_{r \rightarrow 0} p_+(r) = \frac 58$ and $\tilde{p}_+(r=1)=p_+(r=1)=\frac 12$~\footnote{Note that $r=0$ is unphysical since it would not allow substrate cleavage. Therefore, the limit $r \to 0$ has to be understood as a time separation limit where substrate cleavage is much slower than stepping of legs from product sites, \emph{i.e.} $r \ll 1$.}.

There is a special feature of single spiders which makes the definition of $p_+$ straightforward in this case: the spread configuration $ \arraycolsep.5pt \begin{array}{ccc} \bullet & \circ &\bullet \end{array}$ of the spider's legs is unique, since a spider step to the right correspond to a translation of both legs to the right, and hence the configuration before and after a step is the same; cf. \reffig{fig_boundary_periods}\refsfig{fig_boundary_period_single} $(i)$, $(iii)$, and $(vii)$. As we will show below, this is a property which unfortunately does not extend to spider teams. 

A quantity which does not require this uniqueness is the mean number of consecutive directed steps that a spider performs during one boundary period. This quantity will be denoted $\langle S \rangle$ in the following. For single spiders, $\langle S \rangle$ can be calculated as 
\bequ \langle S (p_+) \rangle = \sum_{j=0}^{\infty} j p_j  = \frac{p_+}{1-p_+} \, , \eequ
where \bequ p_j=(p_+)^j (1-p_+) \label{eq_pj} \eequ is the probability that the spider walks precisely $j$ steps during a boundary period, before it leaves the boundary and enters a diffusive period. Let us emphasise that $\langle S \rangle$ is different from the mean \enquote{number of steps the spider makes in the $B$ state}~\cite{Semenov2011}, $\langle S_B \rangle$, as defined by Semenov et al., which counts the number of leg movements (\enquote{leg steps} in our terminology). By contrast, $\langle S \rangle$  only counts a step if \emph{both} legs have been shifted to the right without having moved to the left (\enquote{spider steps}), \emph{i.e.} the number of times the spider consecutively reaches $(vii)$ before $(iii)$, starting from $(i)$ in Fig.~\reffig{fig_boundary_periods}\refsfig{fig_boundary_period_single}. 

The number of consecutive spider steps, $\langle S \rangle$, is equivalent to the number of cleavage events during a boundary period. Not counted is the last cleavage before the spider leaves the boundary period, since this corresponds to a backward step of the spider, cf. Eq.~\eqref{eq_pj}.

\paragraph{Spider teams.}
Clearly, the motion of a single spider is biased only during boundary periods, and undirected during diffusive periods. However, it is manifest that a spider \emph{team}'s motion is not completely diffusive as long as \emph{any of the spiders comprising the team} is in a boundary period. Hence, we consider the spider \emph{team} being in a boundary period if at least one of its spiders resides in a boundary period. In order to compare the performance of individual spiders with that of spider teams, it is now essential to find a way how to count the number of a spider team's steps during a boundary period. Basically, a team moves forward by one step if the boundary between substrate and product sites is shifted forward by one lattice unit on average. To this end we count every cleavage event but for each spider's last cleavage before the \emph{team} leaves the boundary period. In analogy to a single spider, the latter avoids counting those events where the spider team moves away from the boundary and thereby steps backward, cf. Fig. \reffig{fig_boundary_periods}\refsfig{fig_boundary_period_team}.  The number of steps of a spider team is then given by the number of such cleavage events divided by the number of spiders in a team, in analogy to fractional steps of molecular motors like kinesin~\cite{Leduc2007}. For example,
\bequ 
\arraycolsep.5pt \begin{array}{ccccc} \circ & \bullet & \circ & \hat \bullet & \hat \circ \\ \circ & \bullet & \hat  \bullet & \hat \circ & \hat \circ \end{array} \rightarrow \begin{array}{ccccc} \circ & \circ & \bullet & \hat \bullet & \hat \circ \\ \circ & \circ & \bullet & \circ & \hat \bullet \end{array} \label{eq_example_transition}
\eequ 
corresponds to two steps of the lower spider and thus one step for the spider team.

As we consider two or more coupled spiders, the translational symmetry of the state before and after a complete step ($\arraycolsep.5pt \begin{array}{cccc} \bullet & \circ & \hat \bullet & \hat \circ \end{array}$ and $\arraycolsep.5pt \begin{array}{cccc} \circ & \bullet & \circ & \hat \bullet \end{array}$, respectively for a single spider) is broken, likewise the uniqueness of the state which is the first during a boundary period ($\arraycolsep.5pt \begin{array}{ccc} \bullet & \circ & \hat \bullet \end{array}$ for a single spider), is lost. For example,  \bequ \arraycolsep.5pt \begin{array}{ccccc} \bullet & \circ & \hat \bullet & \hat \circ & \hat \circ \\ \circ & \bullet & \bullet & \circ & \hat \circ \end{array} \quad , \quad \begin{array}{ccccc} \circ & \bullet & \circ & \hat \bullet & \hat \circ \\ \circ & \bullet & \bullet & \hat \circ & \hat \circ \end{array} \quad , \quad \begin{array}{ccccc} \circ & \bullet & \circ & \hat \bullet & \hat \circ \\ \circ & \bullet & \circ & \bullet & \hat \circ \end{array} \label{eq_example_entry_states} \eequ all are possible states at the beginning of a boundary period. It is therefore no longer possible to calculate the probability to step to the right (denoted $p_+$ for single spiders) without further specification of these initial states. For spider teams the probability for a forward step explicitly depends on the particular state from which it starts.

This complexity prohibits an analytic treatment of the stochastic dynamics in general. However, if the relative rate of substrate cleavage is small compared to the rate of hopping from product sites, $r \ll 1$, the dynamics become amenable to a theoretical analysis. While in this limit the motion of the boundary between substrate and product sites is slow, the dynamics of spider legs bound to product sites are fast. This suggests to group states into classes characterised by the slow variable, \emph{i.e.} the distance between the ends of the product seas, denoted by $\Delta$. In addition, it turns out to be convenient to introduce subclasses according to the number of spiders attached to substrates, $\sigma$. In the following we will illustrate this for teams comprised of $n=2$ spiders and a leash length $d=2$.  All states  
\bequ \arraycolsep.5pt \begin{array}{ccccc} \circ & \bullet & \circ & \hat \bullet & \hat \circ \\ \circ & \bullet & \circ & \hat \bullet & \hat \circ \end{array} \sim \begin{array}{ccccc} \circ & \circ & \bullet & \hat \bullet & \hat \circ \\ \circ & \bullet & \circ & \hat \bullet & \hat \circ \end{array} \sim \begin{array}{ccccc} \circ & \bullet & \circ & \hat \bullet & \hat \circ \\ \circ & \circ & \bullet & \hat \bullet & \hat \circ \end{array} \sim \begin{array}{ccccc} \circ & \circ & \bullet & \hat \bullet & \hat \circ \\ \circ & \circ & \bullet & \hat \bullet & \hat \circ \end{array} \label{eq_example_eq_class} 
\eequ 
comprise the class
\bequ
\arraycolsep.5pt \biggl[ \begin{array}{ccccc} \circ & \bullet & \circ & \hat \bullet & \hat \circ \\ \circ & \bullet & \circ & \hat \bullet & \hat \circ \end{array} \biggr] =: \bigl[ 0_2 \bigr] = \bigl[ \Delta_\sigma \bigr].\label{eq_def_notation_eq_class} 
\eequ 
Likewise, configurations with $\Delta=0$ and $\sigma=1$, \emph{i.e.} with only one spider having a leg at the boundary, are possible:
\bequ
\arraycolsep.5pt \biggl[ \begin{array}{ccccc} \circ & \bullet & \circ & \hat \bullet & \hat \circ \\ \circ & \bullet & \bullet & \hat \circ & \hat \circ \end{array} \biggr] =: \bigl[ 0_1 \bigr] \label{eq_eq_class_01} .
\eequ 
Here, we made use of the invariance under renumbering of spiders, it is irrelevant if we label the \enquote{upper} spider as 1 and the \enquote{lower} as 2, or the other way round. Hence, irrespective of whether the lower or the upper spider's  leg is bound to a substrate, both contribute to class $\bigl[ 0_1 \bigr]$. That same renumbering symmetry can also be applied when one considers states where the lower and the upper product seas do not end at the same position. This leads to the classes 
\bequ \arraycolsep.5pt \biggl[ \begin{array}{ccccc} \circ & \bullet & \circ & \hat \bullet & \hat \circ \\ \circ & \bullet & \hat \bullet & \hat \circ & \hat \circ \end{array} \biggr]  =: \bigl[ 1_2 \bigr] \quad \textnormal{and} \quad \biggl[ \begin{array}{ccccc} \circ & \bullet & \bullet & \hat \circ & \hat \circ \\ \circ & \bullet & \hat \bullet & \hat \circ & \hat \circ \end{array} \biggr] =: \bigl[ 1_1 \big], \eequ as well as \bequ \arraycolsep.5pt \biggl[ \begin{array}{ccccc} \circ & \circ & \bullet & \bullet & \hat \circ \\ \circ & \bullet & \hat \bullet & \hat \circ & \hat \circ \end{array} \biggr] =: \bigl[ 2_1 \bigr] . \label{eq_eq_class_2_1} 
\eequ  
This completes the list of possible classes with $\sigma \neq 0$ since the constraint on the leash length forbids class $\bigl[ 2_2 \bigr]$, as well as classes $\bigl[ \Delta_\sigma \bigr]$ with $\Delta > 2$. For general $d$,  class $\bigl[ d_2 \bigr]$ and classes with $\Delta>d$ are not allowed.

One can show that the classification of states by means of the distance of the product seas' ends and the number of spiders at the boundary is reflexive, symmetric, and transitive, and hence defines an equivalence relation. Therefore, we tentatively used the symbols $\sim$ and $[ ~ \cdot ~]$ in the previous equations.

Instead of a large number of \enquote{micro}-states, we are now left with only five equivalence classes which include all the spider states at the boundary. The reduction of complexity can be pushed even further: classes $\bigl[ \Delta_1 \bigr]$ with only one leg attached to the substrate are only transient in the sense that they will always decay into classes with two legs attached $\bigl[ \Delta_2 \bigr]$ (as long as $\Delta < d$). Consider, for example, a spider team in class $\bigl[ 0_1 \bigr]$ where one spider's right leg is  attached to a substrate while the other spider's legs are free to move on product sites. Since the diffusion time of legs on products is small compared to the expected residence time $1/r$ of the leg on the substrate, the transition $\bigl[ 0_1 \bigr] \rightarrow \bigl[ 0_2 \bigr]$ is almost certain and happens on a time scale $\sim 1$ (fast compared to substrate cleavage).

All possible transitions between the classes can be visualised as the following reaction scheme:
\bequ 
\begin{array}{ccccccccl}
& & & & & & \bigl[ (d-1)_0 \bigr] & \color{red} \boldsymbol{\rightarrow} & \begin{minipage}{1cm} \textbf{diffusive \\ period} \end{minipage} \\
& & & & & & \color{red} \boldsymbol{\downarrow} & \nwarrow \\
\bigl[0_1\bigr] & & \bigl[1_1\bigr] & & & & \bigl[ (d-1)_1 \bigr]  & \leftarrow & \boldsymbol{\bigl[ d_1 \bigr]} \\ 
\color{red} \boldsymbol{\downarrow} & \crossingarrows  &  \color{red} \boldsymbol{\downarrow} & \crossingarrows  & \dots & \crossingarrows & \color{red} \boldsymbol{\downarrow} & \nearrow & \\ 
\boldsymbol{\bigl[ 0_2 \bigr]} & \rightleftharpoons & \boldsymbol{ \bigl[ 1_2 \bigr]}  & \rightleftharpoons & \mathbf{\dots} & \rightleftharpoons & \boldsymbol{ \bigl[ (d-1)_2 \bigr] }
\end{array} 
\label{eq_reaction_scheme_complex}
\eequ
where $\Delta$ is constant along a column and $\sigma$ along a row, respectively. As explained above, vertical transitions from $\bigl[ \Delta_1 \bigr]$ to $\bigl[ \Delta_2 \bigr]$ are fast (emphasised with bold red/grey arrows in Eq.~\eqref{eq_reaction_scheme_complex}). In contrast, horizontal and diagonal transitions involving substrate cleaving events and hence leading to $\Delta \rightarrow \Delta \pm 1$ are slow. Since vertical transitions occur with certainty and fast, we can eliminate the transient classes $\bigl[ \Delta_1 \bigr]$ and reduce to a reaction scheme for the most stable subclass of each class, shown in boldface in Eq.~\eqref{eq_reaction_scheme_complex} and signified $\bigl[ \Delta \bigr]$ in the following:
\bequ
\arraycolsep.8pt \begin{array}{cccccccccc}
\bigl[ 0 \bigr] & \xrightleftharpoons[\frac 12]{1} & \bigl[ 1 \bigr] & \xrightleftharpoons[\frac 12]{\frac 12} & \dots & \xrightleftharpoons[\frac 12]{\frac 12} & \bigl[ d -1 \bigr] & \xrightleftharpoons[\Pi]{\frac 12} & \bigl[ d \bigr] \xrightarrow{1-\Pi} & \begin{minipage}{1.2cm} \textnormal{diffusive \\ period} \end{minipage} \, .
\end{array} 
\label{eq_reaction_scheme_simple}
\eequ
The numbers above and below the arrows are transition probabilities into the respective classes, reflecting that 
each of the two spiders may cleave a substrate with equal probability for $\Delta < d$. The class $\bigl[ d \bigr] $ has to be treated separately as it constitutes a gate from the boundary into the diffusive period.

Our next set of tasks is now threefold: first, in order for our classification scheme to be a consistent reduction of the stochastic processes, all states comprising the gate class $\bigl[ d \bigr]=\bigl[ d_1 \bigr]$ should have the same \emph{survival probability} $\Pi$, \emph{i.e.} the same probability not to exit into a diffusive period. This is indeed the case for sufficiently small cleavage rates $r$: in the limit $r \to 0$, substrate cleavage events are rare compared to hopping from product sites. Therefore, the dynamics exhibit a time scale separation where all the legs attached to products quickly visit any accessible lattice site while the legs on substrate sites remain stuck. In other words, the dynamics within class $\bigl[ d_1 \bigr]$ are ergodic and equilibrate, and all \enquote{micro}-states effectively reduce to one coarse-grained \enquote{macro}-state, namely the class $\bigl[ d_1 \bigr]$. 
Second, we have to calculate the survival probability $\Pi$ by analysing all the various routes between the \enquote{micro}-states. Third, in order to determine the mean number of consecutive steps $\langle S \rangle$, the reduced reaction scheme of Eq.~\eqref{eq_reaction_scheme_simple} has to be solved.

We now address the calculation of the survival probability $\Pi$. In principle, this can be done for arbitrary complex spider teams. For the purpose of illustration, we continue the example from above with $2$ spiders and a leash length $d=2$. We consider all states comprising class $\bigl[ 2_1 \bigr]$. These are
\begin{eqnarray}\begin{split} & \arraycolsep.5pt  \circled{1} =\begin{array}{ccccc}  \arraycolsep.5pt \circ & \circ & \bullet & \bullet & \hat \circ \\ \circ & \bullet & \hat \bullet & \hat \circ & \hat \circ \end{array} , \circled{2} =\begin{array}{ccccc} \circ & \bullet & \circ & \bullet & \hat \circ \\ \circ & \bullet & \hat \bullet & \hat \circ & \hat \circ \end{array} , \circled{3} =\begin{array}{ccccc} \circ & \bullet & \bullet & \circ & \hat \circ \\ \circ & \bullet & \hat \bullet & \hat \circ & \hat \circ \end{array}, \\
& \arraycolsep.5pt \circled{4} =\begin{array}{ccccc} \bullet & \circ & \bullet & \circ & \hat \circ \\ \circ & \bullet & \hat \bullet & \hat \circ & \hat \circ \end{array}, \circled{5} =\begin{array}{ccccc} \bullet & \bullet & \circ & \circ & \hat \circ \\ \circ & \bullet & \hat \bullet & \hat \circ & \hat \circ \end{array}, \circled{6} =\begin{array}{ccccc} \circ & \bullet & \bullet & \circ & \hat \circ \\ \bullet & \circ & \hat \bullet & \hat \circ & \hat \circ \end{array},  \\
&  \arraycolsep.5pt \circled{7} =\begin{array}{ccccc} \bullet & \circ & \bullet & \circ & \hat \circ \\ \bullet & \circ & \hat \bullet & \hat \circ & \hat \circ \end{array}, \circled{8} =\begin{array}{ccccc} \bullet & \bullet & \circ & \circ & \hat \circ \\ \bullet & \circ & \hat \bullet & \hat \circ & \hat \circ \end{array} , \end{split} \label{eq_sample_states_figure}
\end{eqnarray} 
and their respective \enquote{mirrored} states, \emph{i.e.} the states with spider 1 and 2 interchanged. Let us illustrate the calculation for the particular initial state $\circled{1}$.
Legs unbind from products at rate $1$ and from substrates with rate $r$. 
Hence, from this configuration, the probability that the upper right, or the lower right leg is the first one to unbind is $1/(3+r)$, and $r/(3+r)$, respectively. The left legs unbind first with probability $1/(3+r)$ each. If now, for instance, the lower right leg detaches, it may either reattach to the very same lattice site again, or it may step one site to the right. In either case it cleaves a substrate. Both processes happen with equal probability. Hence, altogether, the transition probability for the lower right leg to step to the right is given by $r/2(3+r)$. The analysis can be continued from the resulting states, and finally leads to a high dimensional system of linear equations. The results obtained by solving the ensuing sets of equations are shown in Fig.~\reffig{fig_justification_eq_class} for all initial states comprising class $\bigl[ d_1 \bigr]$.  
\begin{figure}
\includegraphics{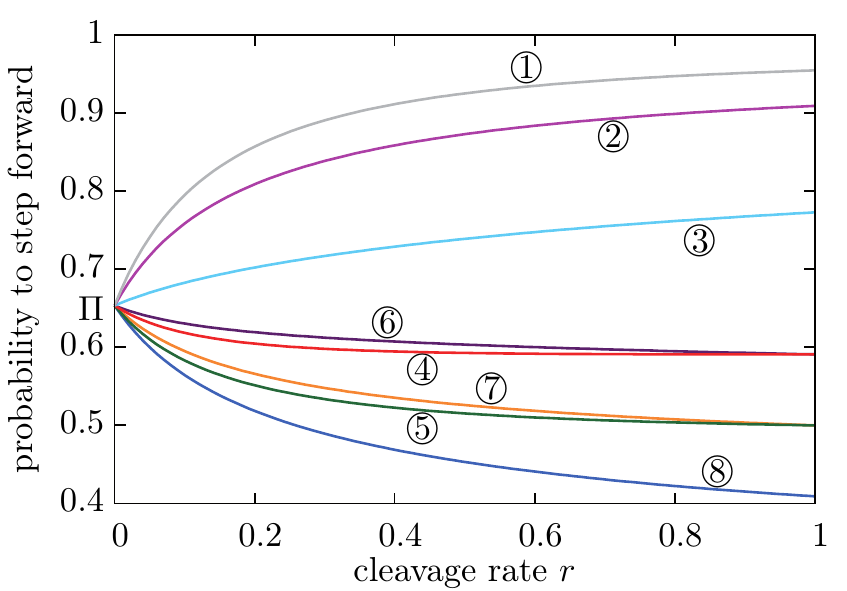}
\caption{\edsug{(Colour online)} Justification for the equivalence classes in the limit $r\to 0$. 
Shown are the analytically calculated probabilities that a spider team ($n=d=2$) successfully completes one step during a boundary period, starting from the specific states \circled{1} -- \circled{8} as given in Eq.~\eqref{eq_sample_states_figure}. Each line corresponds to a state of the equivalence class~$\bigl[ 2_1 \bigr]$ [cf. Eq.~\eqref{eq_eq_class_2_1}]. In the limit $r\to0$, the probability to step forward for all eight states collapses to a fixed value $\Pi\approx 0.65$.\labfig{fig_justification_eq_class} }\end{figure}
Clearly, as $r$ approaches $0$, all survival probabilities approach a single value
\bequ
\Pi=\frac{115}{176} \approx 0.65 \, .
\label{eq:Pi}
\eequ
This result is reassuring, as it confirms our heuristic arguments on the equilibration of states within class $\bigl[ d_1 \bigr]$, and thereby justifies to combine several different states into one class in the limit $r \rightarrow 0$.

All the complexity of calculating the mean number of steps $\langle S \rangle$ of a spider team during a boundary period has now been reduced to analysing the various routes between the \emph{equivalence classes}. Since each transition~\footnote{Following the definition of a step done by a spider team in Sec.~\refsec{sec_boundary_periods}, no step is counted along with the transition $\bigl[ d-1 \bigr] \to \bigl[ d \bigr]$, since this is potentially the last cleavage event of the spider which caused this transition. To compensate this (if this spider makes another cleavage), the transition $\bigl[ d \bigr] \to \bigl[ d-1 \bigr]$ is counted as two (half) steps.
} 
in Eq.~\eqref{eq_reaction_scheme_simple} corresponds to a directed step done during a boundary period, the number of these steps $\langle S \rangle$ is equivalent to the number of (undirected) jumps performed by a simple random walker with reflective, and absorbing boundary conditions on the left, and right end of the reaction scheme, respectively. As detailed in Appendix~\ref{sec_app}, the general solution for the mean number of steps during a boundary period in the limit $r \rightarrow 0$, and for arbitrary $d$, reads 
\bequ 
\langle S(d, r\rightarrow 0) \rangle = \frac{\Pi}{1-\Pi} + (d-1) \frac{1}{1-\Pi} \label{eq_mean_steps} .
\eequ 
For our example of a two-spider team with $d=2$, we obtain using Eq.~\eqref{eq:Pi} 
\bequ
\langle S(r \rightarrow 0)^{n=2}_{d=2} \rangle=\frac{291}{61} \approx 4.77 \, .
\eequ 
We also analysed more complex spider teams with size $n=2,3$ and up to a leash length of $d=4$, and found  even larger mean step numbers, compared to $\frac 53$ for a single spider. 
Obviously, during boundary periods even the simplest spider teams behave significantly more directed and progress a lot further on average, compared to individual spiders. This result is remarkable since directed motion is desirable for applications and a rare feature at the nanoscale.

The analytical results are summarised in Tab.~\ref{tab_results_steps} where they are also compared with Monte Carlo simulations which match them at a very high accuracy. 
\begin{table}[b]
\setlength{\extrarowheight}{2.pt}
\begin{tabular*}{8.6cm}{@{\extracolsep{\fill}}ccc}
\hline \hline
  & $\langle S(r \rightarrow 0)^n_d \rangle$, analytic & $\langle S(r \lesssim 10^{-4})^n_d \rangle$, sim.\\
\hline 
$n=1$ & $\frac 53 \approx 1.6667$ & $1.6672 \pm 0.0015$ \\
$n=2, d=2$ & $\frac{291}{61} \approx 4.770$ & $4.769\pm 0.003$\\
$n=2, d=3$ & $\frac{3170931}{443341} \approx 7.152$ & {$7.146 \pm 0.005 $}\\
$n=2, d=4$ & $\frac{4055316673}{414459263} \approx 9.785$ & $9.785 \pm 0.008$\\
$n=3, d=2$ & $\frac{340881}{48391} \approx 7.044$& $7.042 \pm 0.006$ \\
$n=3, d=3$ & $\frac{16.3745 \dots }{1.34258\dots} \approx 12.196$ & $12.204 \pm 0.012$ \\ \hline \hline
\end{tabular*}
\caption{Comparison of analytic and simulation results for the mean number of steps during a boundary period, $\langle S \rangle$. Analytic values were derived in the limit $r \rightarrow 0$, simulation results were obtained for very small $r\lesssim 10^{-4}$. Simulations and analytical calculations show excellent agreement.
\label{tab_results_steps}}
\end{table}

\subsection{Validity of the equivalence classes\labsec{sec_validity_eq_classes}}
\begin{figure}
\includegraphics{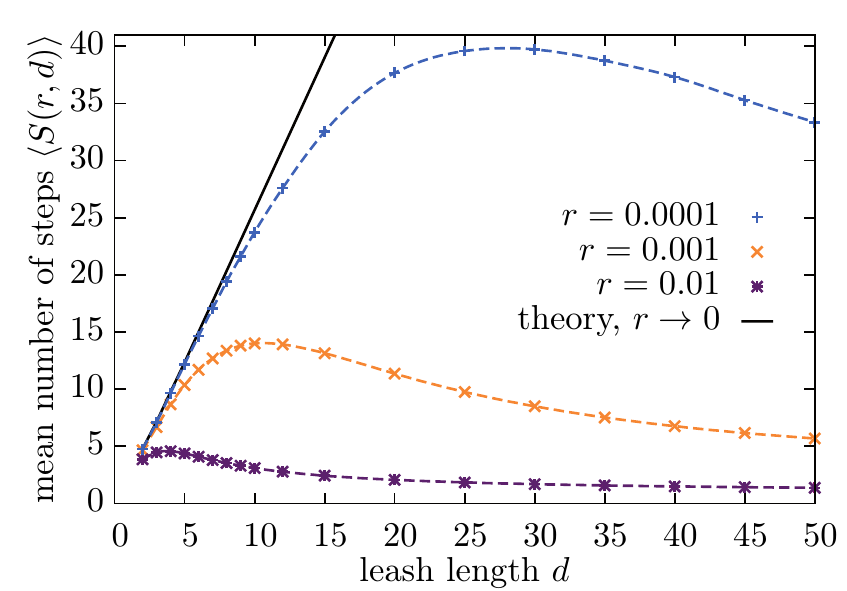}
\caption{(Colour online) Validity of the equivalence class formalism\labfig{fig_validity_eq_classes}. Shown are the simulation results for the mean number of steps, $\langle S \rangle$, for a $2$-spider team and different values of $d$ and cleavage rates $r=0.0001, 0.001, 0.01$; broken lines are a guide to the eye. The theoretical result derived within the equivalence class formalism for $r \rightarrow 0$ (black) is exact for $d=2,3,4$ (Tab.~\ref{tab_results_steps}), and we assumed $\Pi  =\frac 58$ for $d \geq 5$ (Eq.~\eqref{eq_mean_steps}).
}  \end{figure}

With increasing $d$, the spiders forming a team become more and more independent since it is increasingly unlikely that a spider \enquote{feels} the constraint of its teammates. In particular, the probability $\Pi$ that a spider in class $\bigl[ d_1 \bigr]$ reaches $\bigl[ (d-1)_2 \bigr]$ without exiting the boundary period (cf. Eq.~\eqref{eq_reaction_scheme_simple}), converges towards the probability $p_+$ that a single spider makes a step to the right which is $\frac 58$ for $r \rightarrow 0$. Hence, assuming $\Pi=\frac 58$ for large $d$, Eq.~\eqref{eq_mean_steps} would imply that the mean number of steps increases linearly with $d$. Indeed, in the asymptotic limit $r \to 0$ this agrees well with the simulation data. However, with increasing $r$ deviations from this linear behaviour become more and more significant; cf. Fig.~\reffig{fig_validity_eq_classes}. 

This can be explained as follows: for increasing leash length $d$, the configuration space accessible to the spider team becomes progressively larger, so that it takes longer to completely exploit it, \emph{i.e.} the \emph{equilibration time} grows. Conversely, the average time of substrate cleavage scales as $1/r$. With increasing $r$ and/or $d$ these two timescales become comparable. The assumption of time scale separation, on which the reduction of the dynamics to equivalence classes was based on, then becomes invalid. Concluding, the equivalence class concept which we derived in the previous sections provides a very good approximation for small but finite substrate cleavage rates $r$, as long as the leash length $d$ is not too large. 

\subsection{An exact mapping to a confined random walker\labsec{sec_mapping_conf_rw}}

\begin{figure}
\includegraphics[width=8cm]{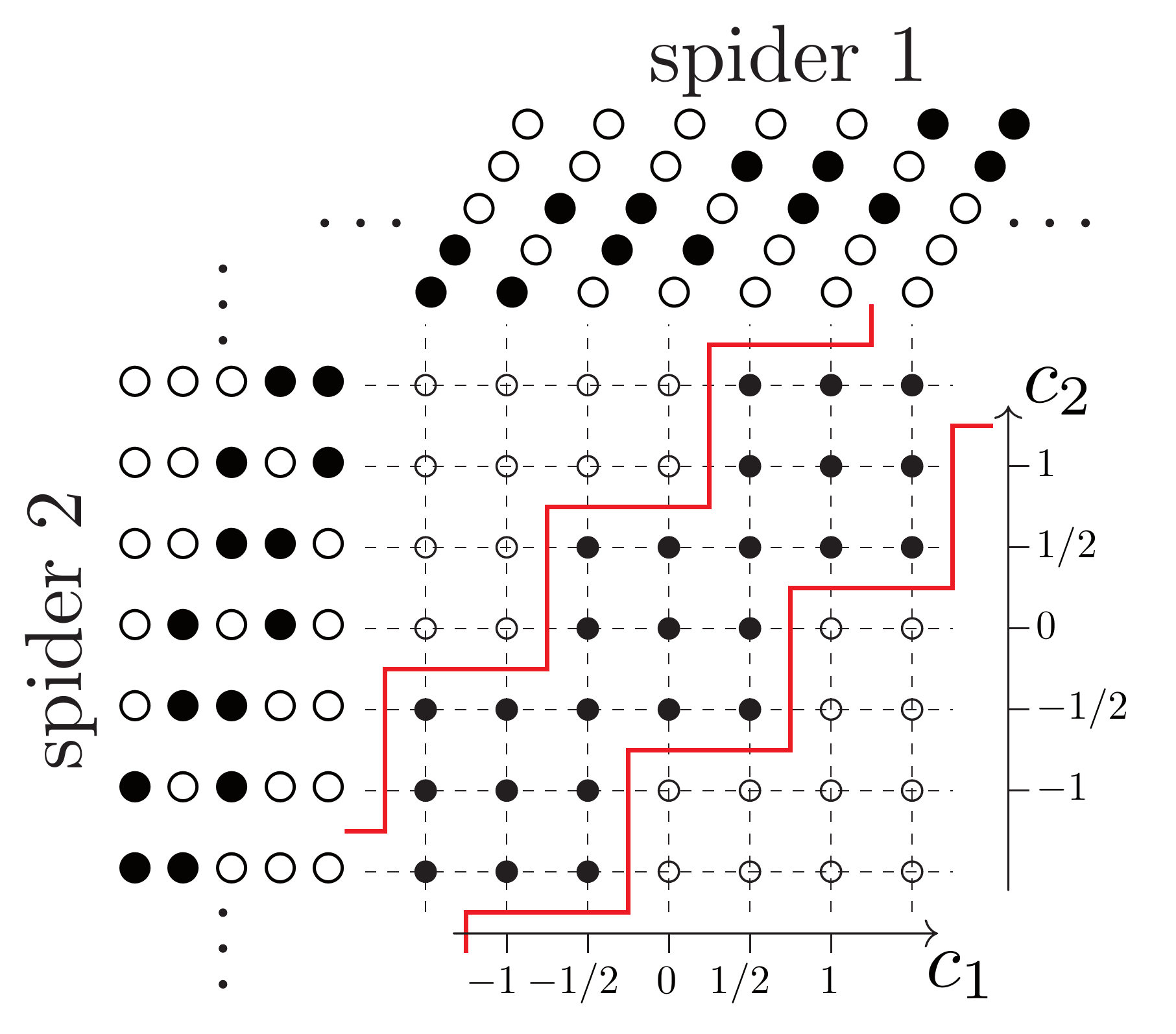}
\caption{\edsug{(Colour online)} A spider team can be mapped to a random walk in a confined environment: transitions of a spider's leg correspond to a change of its centre of mass coordinate $c_i$ of $\pm \frac 12$. Shown is the mapping of a 2-spider team with a leash length $d=2$. The shape of the environment (solid) follows from the leash constraint which confines the span of the spider team. From $d=2$ follows that the leftmost left and the rightmost right leg of the two spiders may be at most two lattice sites apart. With that restriction, the allowed configurations of the team follow directly, as can be seen with some explicit configurations in the left and the top part of the figure.
\labfig{fig_intro_staircase} }
\end{figure}

For a bipedal spider with a maximal span of $s=2$, a single coordinate, the \enquote{centre of mass}-coordinate fully describes the position of the spider's legs. Hence, it is possible to map the motion of the single spider's legs on $\frac 12 \mathbb Z$, the set of integers and half-integers, with hopping of the legs corresponding to changes of the centre of mass. This mapping can be extended for a spider team: the position of a $n$-spider team is characterised by a position on a $n$-dimensional square lattice where each of the $n$ axes corresponds to the centre of mass of one of the spiders comprising the team. The dynamics of a spider team then correspond a trajectory on that lattice. However, due to the leash constraint, not all sites on this lattice are accessible to the spider team. To illustrate this, let us focus on a two-spider team with leash length $d=2$. Fixing the first spider's centre of mass $c_1$, the other spider's centre of mass $c_2$ is restricted to be near $c_1$ due to the leash constraint, cf. Fig.~\reffig{fig_intro_staircase}. We have to distinguish between two cases. Spider 1 is either in a spread or a relaxed configuration, e.g. $c_1=0$ or $c_1=\frac12$, respectively. If it is in the spread configuration $c_1=0$, then the other spider may be in one of three configurations: $c_2 \in \{ -\frac12, 0, \frac12 \}$. For the relaxed configuration $c_1 = \frac12$, there are five configurations possible for the second spider: $-\frac 12, 0, \frac 12, 1,$ and $\frac 32$. Geometrically, this leads to a \emph{staircase} shape for the accessible set of states. For arbitrary $d$, the step width of this staircase generalises to $4d-3$ and $4d-5$. 

While in Sec.~\refsec{sec_diffusive_periods} this mapping will be employed to calculate diffusion constants during diffusive periods, we use it here to illustrate the concept of equivalence classes again. To this end, the mapping is generalised to incorporate substrates as illustrated in shaded colours in Fig.~\reffig{fig_rw_substrate_example_21}: each substrate can be drawn as a box. This is seen as follows: because each spider being at a specific substrate site may either be in a spread or a relaxed configuration, a substrate has to be indicated at \emph{two} different locations in the centre of mass space (thus the width of every box equals $2$). Furthermore, since spider 1 being or not being at a substrate does not affect spider 2, every box indicating a substrate at spider 1's track has to be of a size that it contains all allowed configurations of spider $2$, and vice versa.

We now return to an example discussed in Sec.~\refsec{sec_boundary_periods}: Eq.~\eqref{eq_sample_states_figure} shows all configurations in which spider $1$ has cleaved two more substrates than spider $2$ and only spider $2$ is attached to a substrate. We referred to this set of configurations as the equivalence class $\bigl[ \Delta_\sigma \bigl]=\bigl[ 2_1 \bigl]$. This situation is illustrated in Fig.~\reffig{fig_rw_substrate_example_21}, where there are $\Delta=2$ more boxes (\emph{i.e.} substrates) for spider $2$ than for spider $1$. The eight allowed configurations contained by the ellipse in this figure are only contained in \emph{one} box ($\sigma=1$), such that these states provide an geometrical interpretation of the equivalence class $\bigl[ 2_1 \bigl]$.

Leaving the boundary period in this picture corresponds to removing the encircled box (\emph{i.e.} cleaving the substrate) and stepping down (\emph{i.e.} away from the substrate boundary).
\begin{figure}[htb]
\includegraphics[width=6.8cm]{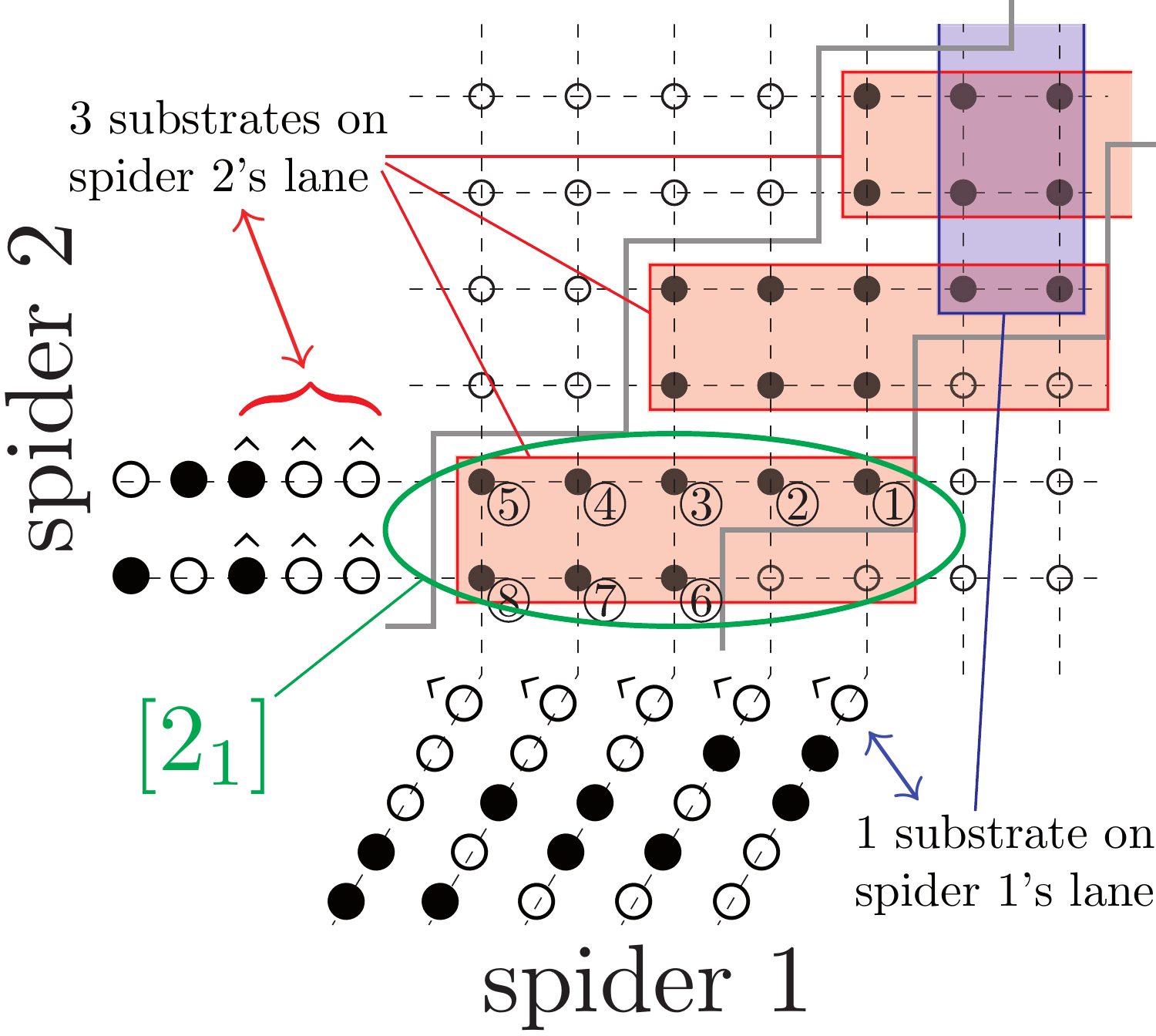}
\caption{\changed{[Changed Figure]} (Colour online) \resug{Substrate in the staircase random walker picture ($n=d=2$ as before). Like in Fig.~\reffig{fig_intro_staircase}, explicit configurations are shown for some points. In addition, boxes are drawn which correspond to the substrates on spider 1's (vertical blue box), or spider 2's (horizontal red boxes) lane. This can be understood as follows: when a spider is attached to a substrate with its right leg, it can be either in the spread or the relaxed configuration. Hence a substrate at position $c$ has to be indicated at \emph{two} points in the centre of mass space, namely at $c-\frac 12$ and $c-1$; therefore the substrate boxes have width $2$. Encircled in the figure are the eight states which have spider $2$ at $\arraycolsep.5pt \begin{array}{ccccc} \bullet & \circ & \hat \bullet & \hat \circ & \hat \circ \end{array}$ or $\arraycolsep.5pt \begin{array}{ccccc} \circ & \bullet & \hat \bullet & \hat \circ & \hat \circ \end{array}$, respectively, and spider $1$ in one of the five states $\arraycolsep.5pt \begin{array}{ccccc} \bullet & \bullet & \circ & \circ & \hat \circ \end{array}, \,\dots , \, \begin{array}{ccccc} \circ & \circ & \bullet & \bullet & \hat \circ \end{array}$. The resulting states correspond clearly to those of Eq.~\eqref{eq_sample_states_figure} and Fig.~\reffig{fig_justification_eq_class}. In the figure, there are three horizontal red boxes (substrates on spider 2's lane), and only one vertical blue box (substrate on lane 1). Hence, the difference of the product sea's ends is $\Delta=2$. Since the encircled states \circled{1} -- \circled{8} have, by direct reading, \emph{only} spider $2$ at a substrate (\emph{i.e.} they are only contained in $\sigma=1$ box), they form the equivalence class $\bigl[\Delta_\sigma \bigr]=\bigl[ 2_1 \bigr]$.} 
\rest{Substrates in the random walker picture. A spider which is at the boundary prevails in one of two configurations (spread or relaxed, respectively). Hence, the existence of a substrate at a site has to be indicated by a box around \emph{two} points in the centre of mass space. Since existence of a substrate on spider 1's lane does not affect spider 2 in any way, a box contains all allowed states of spider 2 (and vice versa). Cleavage of a substrate corresponds to the removal of a box. The different numbers of vertical and horizontal boxes reflect distances of the product seas' ends, denoted by $\Delta$ before; in this figure, $\Delta=2$. The states \circled{1} -- \circled{8} are contained in only one box, corresponding to spider teams with only one spider at a substrate ($\sigma=1$) and hence constitute the equivalence class $\bigl[ 2_1 \bigr]$; cf. Eq.~\eqref{eq_sample_states_figure}. From this class the diffusive period is reached, when a walker removes the encircled box (\emph{i.e.} cleaves the substrate), and subsequently steps down to one of the three points at the bottom.}\labfig{fig_rw_substrate_example_21}}
\end{figure}

\subsection{Diffusive Periods\labsec{sec_diffusive_periods}}
\begin{figure}
\subfigure[\labfig{fig_staircase_twoparts_left}]{\includegraphics{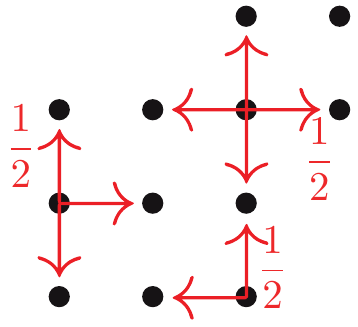}}
\subfigure[\labfig{fig_staircase_twoparts_right}]{\includegraphics{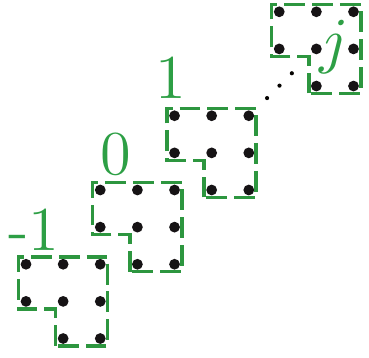}}
\caption{\edsug{(Colour online)} Diffusion in the staircase environment. \refsfig{fig_staircase_twoparts_left} Transition rates between the sites of the staircase environment. Along every arrow drawn, the rate is $\frac 12$ leading to local detailed balance. \refsfig{fig_staircase_twoparts_right} The staircase can be split into elementary cells, numbered with integers.
\labfig{fig_elem_cell}}
\end{figure}

We now employ the mapping of the spider team motion to a confined random walk in order to analyse the spider team's dynamics during a diffusive period. Let us first examine the transition rates between neighbouring points in the confined random walk picture. Consider, for example, the point 
\bequ (c_1,c_2)=(0,0) = \arraycolsep.5pt \begin{array}{ccc} \bullet & \circ & \bullet  \\ \bullet & \circ & \bullet \end{array} \eequ
 in Fig.~\reffig{fig_intro_staircase}. From this configuration, every \emph{leg} may unbind from its product with rate $1$, and then rebind to either the same product site again, or move to the allowed neighbouring site at equal \emph{probability} $\frac 12$. In the confined random walk picture, this leads to transition \emph{rates} of $1 \cdot \frac 12$ along each connection between adjacent sites from $(0,0)$. The same argument applies to any site within the allowed region, so that the transition rate between any two lattice sites equals $\frac 12$, cf. Fig.~\reffig{fig_elem_cell}\refsfig{fig_staircase_twoparts_left}. This leads to the following master equation for the occupation probability $P_{c_1,c_2}$ on the confined lattice:
\bequ
\frac{\d}{\dt} P_{c_1,c_2}=\sum_{\langle c_1,c_2 \rangle} \frac 12  \bigl(P_{\langle c_1,c_2 \rangle}-  P_{c_1,c_2} \bigr) \, , \label{eq_master_equation}
\eequ
where the sum runs over all nearest neighbours $\langle c_1,c_2 \rangle$ of $(c_1,c_2)$. In order to calculate the diffusion coefficient $D=\frac12 \lim_{t\to \infty}\frac{\d}{\dt}\langle x^2(t) \rangle$ we determine the time derivative of the mean square displacement of the spider team:
\bequ \frac{\d}{\d t}\langle x^2(t) \rangle=  \sum_{(c_1,c_2) \in \mathcal C} x^2_{c_1,c_2} 
\sum_{\langle c_1,c_2 \rangle} \frac 12 \bigl(P_{\langle c_1,c_2 \rangle}-  P_{c_1,c_2} \bigr) ,
\label{eq_def_msd}\eequ 
where $x_{c_1,c_2}=\frac 12 (c_1+c_2)$ is the position of the spider team on the molecular track for given values of $c_1$ and $c_2$, and the summation extends over all $(c_1,c_2)$ within the allowed region $\mathcal{C}$.
This equation can be reorganised such that
\bequ
\frac{\d}{\d t} \langle x^2(t) \rangle = \sum_{\mathcal C} P_{c_1,c_2} \sum_{\langle c_1,c_2 \rangle} \frac 12 \bigl( x^2_{\langle c_1,c_2 \rangle}-x^2_{c_1,c_2} \bigr) \, .
\eequ

To evaluate this expression we split the lattice into elementary cells as shown in Fig.~\reffig{fig_elem_cell}\refsfig{fig_staircase_twoparts_right}, and use that for asymptotically large times, $t \to \infty$, the probability density $P$ varies only \rest{slowly along the diagonal $c_1=c_2$} \resug{little between neighbouring elementary cells. This follows from translational symmetry; every cell obeys the same master equation}.  The master  equation, Eq.~\eqref{eq_master_equation}, then implies a nearly uniform probability distribution within each elementary cell $j$~\footnote{Note that global equilibrium is never reached in this system due to the open boundaries.}. Upon assuming a constant value $P_j$ within each unit cell, carrying out the sum over an arbitrary elementary cell $j$ leads to a further simplification
\bequ
\sum_{\mathcal C_j} P_j \sum_{\langle c_1,c_2 \rangle} \frac 12 \bigl( x^2_{\langle c_1,c_2 \rangle}-x^2_{c_1,c_2} \bigr) = \frac 12 P_j \, ,
\eequ
independent of $j$. Altogether, we obtain
\begin{eqnarray}\begin{split}
 \lim_{t \to \infty} \frac{\d}{\dt} &\langle  x^2(t) \rangle \approx \sum_{j=-\infty}^{\infty} \frac{1}{2} P_j \stackrel{(*)}{\approx} \sum_{j=-\infty}^{\infty} \sum_{\mathcal C_j} \frac{\frac 12}{8} P_{c_1,c_2} \\ &= \frac{1}{16} \sum_{\mathcal C} P_{c_1,c_2} \stackrel{(\dagger)}{=} \frac{1}{16} = 2 D\, ,
 \end{split}\end{eqnarray}
where in $(*)$ we used that each elementary cell comprises $8$ points, and in $(\dagger)$ we employed the normalisation condition for $P$. 
This procedure can be generalised for arbitrary $d$. The formula for the diffusion constants for $n=2$ then reads  \bequ D(d)=\frac{1}{16}+\frac{1}{32(1-d)} \label{eq_diff_n2}\, .\eequ

This theoretical result agrees well with simulation data for the diffusion constant $D$, as a function of the leash length $d$, see Fig.~\reffig{fig_diffusion_constants}. 

\begin{figure}
\includegraphics{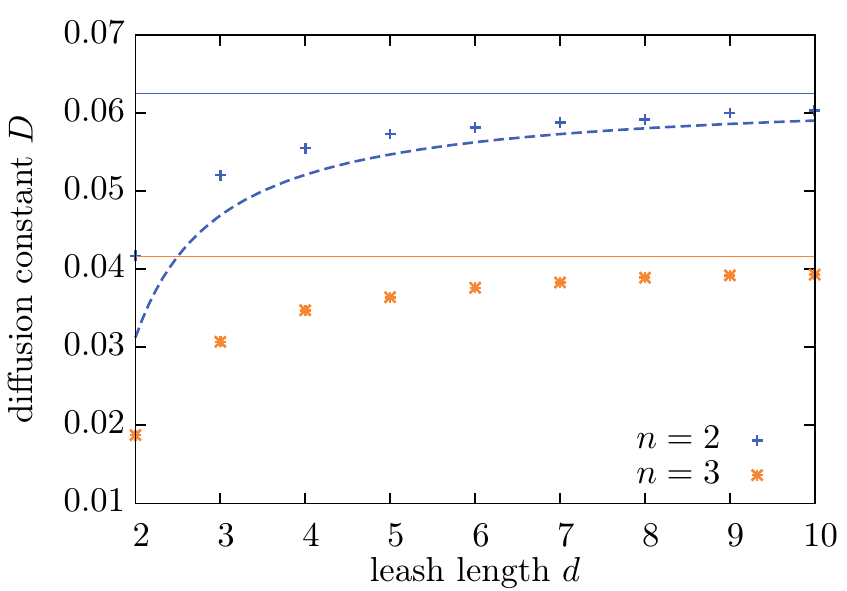}
\caption{(Colour online) Diffusion constants as a function of the leash length $d$ for $n=2$ and $3$ spiders. The dashed line shows the theoretical result for $n=2$ (Eq.~\eqref{eq_diff_n2}); solid lines are asymptotics for $d\to\infty$. Our theoretical approximation is in good agreement with  simulation data (points).\labfig{fig_diffusion_constants} }
\end{figure}

\section{Optimisation of directed motion\labsec{optimisation}}

In the previous sections we mainly focussed on ensemble properties of spider teams. However, in experiments or applications one has to deal with single realisations of the stochastic process, \emph{i.e.} single trajectories, cf. Fig.~\reffig{speed}\refsfig{fig_sample_trajectory}. Since it is desirable to achieve a molecular motor design that works reliably, one would like to minimise the randomness of the trajectory, \emph{i.e.} the motion's standard deviation 
\bequ
\sigma=\sqrt{\big\langle (x -\langle x\rangle)^2\big\rangle}\, .
\eequ
It is interesting to ask how the microscopic properties of the spider team $(n,d)$ influence $\sigma$: can we optimise the performance of a spider team? Is there an optimal choice of parameters, $n$ and $d$, which reduces the randomness of a spider teams' motion to a minimum? 

The randomness is determined by the interplay between the dynamics of the spider team during its different episodes of motion, \emph{i.e.} the boundary periods and the diffusive periods. For each episode we found a characteristic feature: during boundary periods the spider team motion is essentially ballistic which can be quantified in terms of the mean number of consecutive steps $\langle S\rangle$ (cf. Eq.~\eqref{eq_mean_steps}). In contrast, during a diffusive period the spider team performs a random walk with a diffusion constant $D$ (cf. Eq.~\eqref{eq_diff_n2}).

We have already learned in Sec.~\refsec{sec_validity_eq_classes} and Fig.~\reffig{fig_validity_eq_classes} that there is an optimal choice of parameters for the number of consecutive directed steps during a boundary period (see Fig.~\reffig{fig_validity_eq_classes}). One could now na\"ively conclude that the predictability of a spider team motion can as well be optimised with the same set of parameters. However, this argument would overlook the impact of the diffusive periods. Indeed, there are several effects which influence the randomness during these episodes:
\begin{enumerate}[(i)]
\item \label{eff_i}In Sec.~\refsec{sec_diffusive_periods} we noted that the diffusion constant $D$ grows with the leash length $d$ (Eq.~\eqref{eq_diff_n2} and Fig.~\reffig{fig_diffusion_constants}). Since $D$ determines the mean square displacement during a diffusive period, increasing $d$ would then also imply a \emph{greater} randomness,~$\sigma$.
\item \label{eff_ii}Conversely, a higher diffusion coefficient during diffusive periods speeds up all dynamic processes. Thus, in a given time window, larger $d$ make it more probable for a spider team to return to the boundary and start moving ballistically~\footnote{Note that although the random walker is recurrent its recurrence time is infinite~\cite{Polya1921}.}.
\end{enumerate}
The combined effect of these two processes can be estimated by analysing a random walker with an absorbing boundary. In one dimension, one finds that $\langle x^2 (t)\rangle\propto \sqrt{Dt}$~\cite{vanKampen2007,Redner2009}.  Hence, (\ref{eff_i}) and (\ref{eff_ii}) together would lead to an \emph{increase} of $\sigma$ with $d$.
\begin{enumerate}[(i)]\setcounter{enumi}{2}
\item \label{eff_iii}Consider the geometrical interpretation of the transition from the boundary period to the diffusive period as given in Fig.~\reffig{fig_rw_substrate_example_21}. In this picture, entering a diffusive period corresponds to removing the lowermost red box, and stepping to one of the three points on the very bottom. Right after this transition, the average minimal distance $\langle{x_0}\rangle$ of the spider team from the boundary is therefore given by \bequ \langle{x_0}\rangle=\frac 14\left( 3+\frac{3}{4d-5}\right)\, ,\eequ
as can be inferred from counting the different transition pathways. Hence, with increasing $d$, the spider team entering the diffusive period remains closer to the boundary, and is thereby more likely to reenter a boundary period quickly.
\item \label{eff_iv} In Sec.~\refsec{sec_mapping_conf_rw} we have shown that with increasing leash length $d$ the number of pathways in state space to reenter a boundary period also increases. Pictorially, this can be inferred from the mapping of the spider team's motion to a random walker in a staircase environment: the longer the leash length $d$ the larger is the \enquote{angle} under which a random walker sees the boundary of the staircase. Thus, when the random walker takes an arbitrary direction the probability that it walks toward the boundary is increasing with $d$. \end{enumerate}

Since their is no unique trend in the various effects discussed above (\ref{eff_i})-(\ref{eff_iv}), it is difficult to conclude what would be the dominant effect of the diffusive period on the randomness. Therefore, we numerically determine the randomness of the spider team during diffusive periods~\footnote{It is convenient to run the simulations in the staircase picture (cf. Fig.~\reffig{fig_rw_substrate_example_21}). We assume that each of the points at the very bottom of this figure is an equally likely starting point, and set up an absorbing boundary at the lowermost substrate box.}; 
this quantity is depicted in Fig.~\reffig{fig_trend_inverse}. We observe that the mean squared distance from the boundary is smaller for larger $d$ at all times. This implies that --- considering only diffusive periods --- increasing $d$ leads to a \emph{reduction} of the randomness. From this we can infer that the effects (\ref{eff_iii}) and (\ref{eff_iv}), which decrease the randomness of the process with increasing $d$, overcompensate the effects (\ref{eff_i}) and (\ref{eff_ii}).

\begin{figure}
\includegraphics{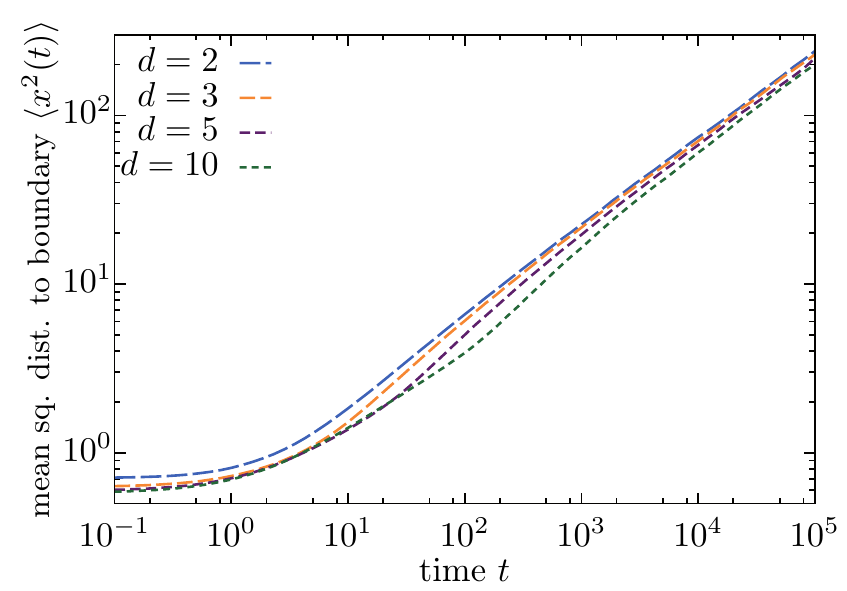}
\caption{(Colour online) Randomness during a diffusive period. Shown is the mean squared minimal distance to the boundary of a random walker in the staircase environment \authsug{($n=2$)}, Fig.~\reffig{fig_intro_staircase}. The walkers start randomly along every point which provides an entrance to the diffusive period (for $d=2$, these are the three points at the very bottom of Fig.~\reffig{fig_rw_substrate_example_21}); they are absorbed when they reach the boundary (which is the second substrate box in Fig.~\reffig{fig_rw_substrate_example_21}; note that the lowermost box has been removed when the walker entered the diffusive period!). Obviously, the mean squared distance is the greater the smaller $d$ is. Increasing $d$ thus decreases the randomness.\labfig{fig_trend_inverse}}
\end{figure}

Altogether we can now conclude the influence of the diffusive periods as follows:
\bequ
d \nearrow \quad \Rightarrow \quad \sigma \searrow\, .\nonumber
\eequ 
Analogously we can decipher the influence of boundary periods.
Going back to Fig.~\reffig{fig_validity_eq_classes} we observe: 

\bequ
\begin{array}{cc}
d<d^{\textnormal{opt}}_{\langle S \rangle}: & d \nearrow \quad \Rightarrow \quad \langle S \rangle  \nearrow \quad \Rightarrow  \quad \sigma \searrow   \, ,\vspace{.2cm} \\ 
d>d^{\textnormal{opt}}_{\langle S \rangle}: & d \nearrow \quad \Rightarrow \quad \langle S \rangle  \searrow \quad \Rightarrow  \quad \sigma \nearrow \, . \end{array}\nonumber
\eequ

These considerations explain that \emph{if} there is an optimal value $d^\textnormal{opt}_\sigma$, it must be found beyond $d^{\textnormal{opt}}_{\langle S \rangle}$. This is in agreement with our data:
Figure~\reffig{fig_comparison_optimal_d} shows the existence of a minimum of the randomness, and its positioning with respect to $d^\textnormal{opt}_{\langle S \rangle}$. Remarkably, the positions of both optima are strongly correlated (see Tab.~\ref{tab_optimal_values}).  

In conclusion, our analysis shows that the randomness of the spider team is mainly determined by the mean number of steps $\langle S\rangle$ during boundary periods. Diffusive periods have only a small effect on the randomness and change the optimal parameters only slightly.

\begin{figure}
\includegraphics[width=1\columnwidth]{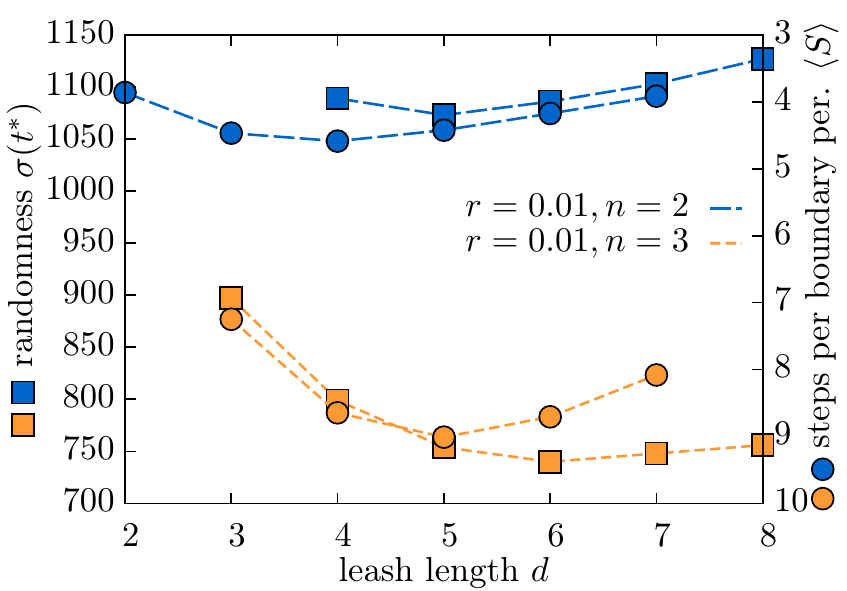}
\caption{\changed{[Changed symbols]} (Colour online) Standard deviation $\sigma$ of the spiders movement and the mean number of steps $\langle S \rangle$ as a function of the leash length $d$. Both $\sigma$ and $\langle S \rangle$ show extrema. To emphasise the correspondence between the minimum of $\sigma$ and the maximum of $\langle S \rangle$ (cf. Fig.~\reffig{fig_validity_eq_classes}), the $\langle S \rangle$-axis is drawn in reverse (see right scale). 
$\sigma$ is measured at the time $t^*$ when the mean displacement $\langle x \rangle$ equals 1000. This choice is arbitrary; for smaller values the minima of $\sigma$ persist, but are less pronounced, cf. Fig.~\reffig{properties}\refsfig{fig_mean_displacement}.
\labfig{fig_comparison_optimal_d}
} 
\end{figure}

\begin{table}[b]
\begin{tabular*}{8.6cm}{@{\extracolsep{\fill}}rp{1cm}ccp{1.5cm}cc}
&&\multicolumn{2}{c}{$n=2$}&&\multicolumn{2}{c}{$n=3$} \\
\hline \hline
$r$ && $d^{\textnormal{opt}}_{\langle S \rangle}$ & $d^{\textnormal{opt}}_{\sigma}$ & &$d^{\textnormal{opt}}_{\langle S \rangle}$ & $d^{\textnormal{opt}}_{\sigma}$\\
\hline
0.001 && $\sim$10-11 & $\sim$13 \\
0.01 && 4 & 5 && 5 & 6\\
0.02 && 3 & 4 && 3-4 & 4-5\\
0.05 && 2 & 3 \\
0.1 && 2 & 3 && 2 & 3\\
0.2 && 2 & 2 && 2 & 3\\
\hline \hline
\end{tabular*}
\caption{\label{tab_optimal_values}Optimal values of $d$ for $n=2$ and $3$, and several values of $r$. Compared are the values of $d$ which maximise the mean number of steps during a boundary period, and that which minimise the randomness (for a comparison see Fig.~\reffig{fig_comparison_optimal_d}). Clearly, both values of $d$ are closely correlated, where $d^{\textnormal{opt}}_{\langle S \rangle}$ is only slightly smaller than $d^{\textnormal{opt}}_{\sigma}$.
}
\end{table}

\section{Conclusions\labsec{conclusions}}

Based on existing models for molecular spiders~\cite{Antal2007,Antal2007a}, we proposed a model for a spider team that explores the collective behaviour of cooperating spiders: in our model, bipedal spiders are jointly attached to a (zero-mass) linking cargo. Each spider walks on its own one-dimensional track. This leads to a spacial constraint which can be characterised by the maximal span $d$, of the resulting spider team.

Depending on the cleavage rate of the substrate $r<1$, the number of coupled spiders $n$, and the leash length $d$, we found that the coupling leads to a significant enhancement of many of the spider's motor properties: spider teams show a significant increase of their mean displacement; their motion is a lot less random; the ensemble's velocity can be increased by more than an order of magnitude; and the superdiffusive behaviour lasts longer for orders of magnitude in time. 
Unlike single spiders, cooperating spiders could therefore --- at least in theory --- be employed for executing well-defined tasks reliably.

Like their individual counterparts~\cite{Semenov2011}, spider teams' motion can be characterised as being in either a boundary, or a diffusive period. 
We found that the characteristic quantity is the mean number of consecutive directed steps, $\langle S \rangle$, which a spider team performs during a boundary period. In simplified language, $\langle S \rangle$ integrates the number of steps which the spider teams walks, as long as it stays in the vicinity of the co-moving boundary between substrate and product sites. $\langle S \rangle$ is closely related to the bias $p_+$ of single spiders~\cite{Antal2007}.
For small $r$, we succeeded in calculating $\langle S \rangle$ analytically through an equivalence class formalism which made use of the time scale separation of dwell times on products and substrates. This formalism is exact for $r \rightarrow 0$, regardless of the number of coupled spiders and the tightness of the coupling. We explicitly calculated values for various small spider teams, and find excellent agreement with simulation data. For small but finite cleavage rates $r$, the formalism still holds as an approximation for relatively tight coupling. We found that in this case there is an optimal value for the coupling tightness $d$ which maximises the mean number of steps.

Next, we provided a mapping of the stochastic motion of an $n$-spider team to a random walker in an $n$-dimensional environment. The motion is confined between two parallel boundaries which have the shape of staircases. This mapping is exact and allows a complementary interpretation for the equivalence classes: substrates can be drawn as boxes which are easy to enter for random walkers but impossible to leave without removing, which happens slowly on a timescale $r^{-1}$. It is then straightforward to see that an equivalence class corresponds to an intersection of boxes (cf. Fig.~\reffig{fig_intro_staircase}). The staircase picture also allows to quantify the dynamics during the diffusive periods of spider teams: in that case, boxes can be ignored and spider teams correspond to ordinary diffusive random walkers on the confined lattice. We calculated the diffusion constants for 2-spider teams and find good agreement with simulation data.

The analysis of the mean number of consecutive steps during a boundary period, $\langle S \rangle$ (which shows a maximum for some value of the leash length $d$),
taken together with the diffusion constants $D$ (which grow with $d$) allow for a comprehensive explanation of our observations. We show that the optimal value of  $d$ that minimises the randomness (which involves boundary and diffusive periods) differs only slightly from the leash length maximising the mean number of steps during a boundary period; see Fig.~\reffig{fig_comparison_optimal_d}.

The staircase picture also illustrates that despite the difference in complexity, a single spider and a spider team can both be described by similar effective random walk models: the motion of a bipedal spider which has a non-trivial stepping gait can be fully described by its centre of mass coordinate which performs simple one-dimensional random walks~\cite{Antal2007}. Likewise, the motion of an $n$-spider team which involves complicated interactions between the spiders can equivalently be described by another single coordinate which performs $n$-dimensional random walks that are however geometrically confined due to the leash constraint.

Our results show that the primary factor for improving the motor properties of molecular spiders is the accessibility of substrate sites for the spider legs: while single spiders only have access to one substrate at a time, an $n$-spider team can reach $n$ substrates. This would imply that there is a significant difference between truly one-dimensional spiders~\cite{Antal2007} and quasi one-dimensional spiders~\cite{Lund2010}.  This is enforced by a very recent study of Olah et al.~\cite{Olah2012} who examined molecular spiders on a narrow $2$-dimensional lattice.  As well, it is in full accordance with recent data by Samii et al.~\cite{Samii2011} who concentrated on hand-over-hand spiders: they showed that motor properties of this class of spiders which have access to more than one substrate site at a time are superior to inchworm spiders which can only reach one substrate at once~\cite{Samii2010,Samii2011}. 

The results presented here can be extended in multiple ways. In analogy to individual spiders, further studies could concentrate on varying design specifics like the number or the length of legs~\cite{Samii2011}. Likewise, the underlying chemical processes~\cite{Pei2006,Lund2010} could be modelled in greater molecular detail also for spider teams. Similarly, the team's spiders' stepping gait could be varied, potentially profiting from studies about the motion of individual hand-over-hand spiders with more than two legs~\cite{Samii2011} which seem to be difficult to realise in the experiment. 

Unlike other studies (e.g.~\cite{Semenov2011}) which have extensively investigated the role of the cleavage rate $r$, our focus was different and the variation of $r$ was only a side aspect of this work. Nevertheless, our analysis hints towards a scaling behaviour which maps the quantity $\langle S(r,d)\rangle$ to a universal form $\tilde S(\tilde d)$ which is independent of $r$.
In this spirit, it would also be interesting to study the connection of the optimal leash length and the cleavage rate $r$. It appears that this relation might be rather simple for a wide parameter range, although its mathematical formulation seems to be very complex. The difficulty is that the simplified formulation of the problem presented here, \emph{i.e.} the equivalence classes, can not be applied directly. One possibility to address this problem might lie in drawing analogies from related models like the burnt-bridge model~\cite{Antal2005}. For example, it has been studied for dimeric motor molecules~\cite{Morozov2007} and as an exclusion process~\cite{Schulz2011}.

Our results might also be relevant to study collective properties of molecular motor assemblies theoretically, cf. Ref.~\cite{Guerin2010} and the references therein. These models are relevant to understand the interplay between biological motor molecules like kinesin, dynein and myosin inside cells~\cite{Gross2007,Holzbaur2010}. In contrast to spiders, biological motors are fuelled by ATP hydrolysis; they can build up significant pulling forces due to strong mechanochemical coupling~\cite{Kolomeisky2007}. In particular, recent experiments addressed the complex interplay of multiple coupled kinesin motor proteins where the motors are coupled via a DNA leash of certain length. It is interesting to note how in these experiments teams of two kinesin motors outperform a single motor in terms of run-length and pulling forces~\cite{Rogers2009,Driver2011,Jamison2012}. \authsug{Similarly, cooperative effects also improve the properties of two coupled burnt-bridge motors modeling collagenase transport\cite{Morozov2007}.}

In conclusion, we believe that our model of coupling molecular spiders provides insight on how cooperative behaviour evolves on the molecular scale. We hope that our ideas about molecular spiders help advance a young and fast growing field in which much focus is put on the construction of novel, more efficient, molecular designs~\cite{Pinheiro2011}. We believe that our findings are not limited to the case of molecular spiders, but apply to molecular machines working together in general.

\begin{acknowledgments}
We thank Anatoly Kolomeisky for fruitful discussions \authsug{and comments on the manuscript}. This project was supported by the the German Excellence Initiative via the program ``Nanosystems Initiative Munich'' (NIM). M. R. gratefully acknowledges a scholarship by the Cusanuswerk.
\end{acknowledgments}

\appendix
\section{Derivation of Eq.~\eqref{eq_mean_steps}\label{sec_app}}
We analyse the graph for a $2$-spider team with arbitrary $d$ as depicted in Eq.~\eqref{eq_reaction_scheme_simple}.
According to this graph, transitions $\bigl[ i \bigr] \to \bigl[ i \pm 1 \bigr]$ are equally likely as long as $i<d$, whereas $\bigl[ d \bigr] \to \bigl[ d-1 \bigr]$ happens at probability $\Pi$. During every transition, the spider team performs a fractional step $\frac 1n=\frac 12$. Only during the transition $\bigl[ d \bigr] \to \bigl[ d-1 \bigr]$, no step is integrated; in return, $\bigl[ d-1 \bigr] \to \bigl[ d \bigr]$ leads to a whole step for the team. This is due to the very definition of the number of steps during a boundary period, which comprises all cleavages but for each spider's last cleavage before the team enters the diffusive period.

With these preparations, we can now establish the probabilities $p(j \vert \bigl[ i \bigr])$ that a spider team, being in class $\bigl[ i \bigr]$, performs exactly $j$ steps before leaving into the diffusive period. These read
\begin{eqnarray}\begin{split}
& p(j \vert [0]) = p(j-\frac 12 \vert [1]) \, ,\\
& p(j \vert [i]) = \frac 12 \Bigl( p(j\!-\!\frac 12 \vert [i-1]) + p(j\!-\!\frac 12 \vert [i+1]) \Bigr) \, ,\\
& p(j \vert [d-1])  =\frac 12 \Bigl( p( j-\frac 12 \vert [d-2])+ p(j \vert [d]) \Bigr) \, , \\
& p(j \vert [d]) = \Pi p(j-1 \vert [d-1]) \, , \label{eq_app_pj}
\end{split}\end{eqnarray}
where $0<i<d-1$. The \emph{mean} number of steps $\langle S(x) \rangle$ which a spider team walks from class $\bigl[ x \bigr]$ until going to the diffusive period is then given by
\bequ \langle S(x) \rangle = \sum_{j=0,\frac 12,\dots}^{\infty} j p(j \vert [x]). \label{eq_app_S}\eequ

Inserting Eq.~\eqref{eq_app_S} into Eq.~\eqref{eq_app_pj}, and by renumbering indexes we obtain
\begin{eqnarray}\begin{split} 
&\langle S(0) \rangle = \frac 12 + \langle S(1) \rangle \, , \\
&\langle S(i) \rangle = \frac 12 + \frac 12 \langle S(i-1) \rangle + \frac 12 \langle S(i+1) \rangle, \\ 
&\langle S(d-1) \rangle = \frac 14 + \frac 12 \langle S(d-2) \rangle + \frac 12 \langle S(d) \rangle , \\ 
&\langle S(d) \rangle = \Pi + \Pi \langle S(d-1) \rangle \, , \end{split}\end{eqnarray} 
where again $0<i<d-1$.
Solving this system of equations, we obtain the recursion relation \bequ \langle S(k) \rangle=\langle S(k+1) \rangle +k+\frac 12 \eequ for $0 \leq k<d-2$. Substituting this into the remaining equations leads to \bequ \langle S(d) \rangle = \frac{\Pi d}{1-\Pi} \, ,\eequ 
and finally 
\bequ \langle S(d-1) \rangle = \frac{d}{1-\Pi} - 1 = \bigl( d-1 \bigr) \frac{1}{1-\Pi} + \frac{\Pi}{1-\Pi}\, . \eequ Since a spider always enters a boundary period in class $\bigl[ d-1 \bigr]$ in the limit $r \to 0$ (cf. Eq.~\eqref{eq_reaction_scheme_complex}), the last equation is equivalent to $\langle S \rangle$, Eq.~\eqref{eq_mean_steps}.


%

\bibliographystyle{apsrev4-1}

\end{document}